\pdfoutput=1 

\documentclass[a4paper,11pt]{article}

\usepackage{jheppub} 

\usepackage[T1]{fontenc} 
\usepackage{braket}
\usepackage{siunitx}

\title{\boldmath Quantum-correlated measurements of $D\to K^{0}_{\rm S}\pi^{+}\pi^{-⁠}\pi^{0}$ decays and consequences for the determination of the CKM angle $\gamma$}

\author[a,1]{P. K. Resmi,\note{Corresponding author.}}
\author[a]{J. Libby,}
\author[b]{S. Malde}
\author[b]{and G. Wilkinson}

\affiliation[a]{Indian Institute of Technology Madras,\\Chennai 600036, India}
\affiliation[b]{University of Oxford,\\ Denys Wilkinson Building, Keble Road, OX1 3RH, United Kingdom}

\emailAdd{resmipk@physics.iitm.ac.in}
\emailAdd{libby@iitm.ac.in}
\emailAdd{Sneha.Malde@physics.ox.ac.uk}
\emailAdd{Guy.Wilkinson@cern.ch}

\abstract{We perform quantum-correlated measurements of the decay $D~\to~K^{0}_{\rm S}\pi^{+}\pi^{-⁠}\pi^{0}$ using a data sample corresponding to an integrated luminosity of 0.82 fb$^{-1}$ collected at the $\psi(3770)$ resonance by the CLEO-c detector. The value of the $CP$-even fraction $F_{+}$ is determined to be 0.238~$\pm$~0.012~$\pm$~0.012. The strong-phase differences are also measured in different regions of $K^{0}_{\rm S}\pi^{+}\pi^{-⁠}\pi^{0}$ phase space by binning around the intermediate resonances present. The potential sensitivity of the results for determining the CKM angle $\gamma$ from $B^{\pm}~\to~D(K^{0}_{\rm S}\pi^{+}\pi^{-}\pi^{0})K^{\pm}$ decays is also discussed. }


\begin{document} 
\maketitle
\flushbottom

\section{Introduction}
\label{sec:intro}

The CKM~\cite{C,KM} angle $\gamma \equiv \rm{arg}(-V_{ud}V_{ub}^{*}/V_{cd}V_{cb}^{*})$, sometimes denoted as $\phi_3$, can be measured using decays $B^{\pm}~\to~DK^{\pm}$ with $D$ being a neutral charm meson reconstructed in a final state common to both $D^{0}$ and $\bar{D^{0}}$ decays. The current uncertainty on $\gamma$ is significantly larger than that of the Standard Model (SM) prediction, which is calculated from the assumption of unitarity and measurements of other parameters in the CKM matrix~\cite{PDG}. This is due to the small branching fraction of decays sensitive to $\gamma$. A more precise measurement of $\gamma$ is crucial for testing the SM description of $CP$ violation and probing for new physics effects. The data collected at detectors such as BABAR, Belle, LHCb or the future Belle II experiment can be used to determine $\gamma$. The $D$ decay final states so far studied include those that are either $CP$-eigenstates, flavour specific or self-conjugate. The statistical uncertainty on $\gamma$ can be reduced if information from additional $D$ meson final states is included, which in practice means new three and four-body decay modes.  However, the use of multibody final states requires knowledge of the strong-phase difference between the $D^0$ and $\bar{D}^{0}$ that varies over the phase space. Determining the fractional $CP$ content of the $D$ meson final state is also helpful. The required information can be obtained by studying quantum-correlated $D\bar{D}$ mesons produced in $e^{+}e^{-}$ collisions at an energy corresponding to the mass of the $\psi(3770)$. 

Here, we present the first quantum-correlated analysis of the decay $D\to K_{\rm S}^{0}\pi^{+}\pi^{-}\pi^{0}$, which has a branching fraction of 5.2$\%$~\cite{PDG}, which is large compared to that of other multibody final states. The study is made with data collected by the CLEO-c detector, corresponding to an integrated luminosity of 0.82~fb$^{-1}$. We determine the $CP$-even fraction $F_{+}$ of the decay which makes it potentially useful in a quasi-GLW~\cite{GLW1, GLW2} analysis along with other $CP$ eigenstates~\cite{MNayak}. Furthermore, this multibody self-conjugate decay occurs via many intermediate resonances, such as $K_{\rm S}^{0}\omega$ and $K^{*\pm}\rho^{\mp}$, hence if the strong-phase difference variation over the phase space is known, a GGSZ-style~\cite{GGSZ, GGSZ2} analysis to determine $\gamma$ from this final state alone is possible. Hence measurements of the relative strong-phase are performed in localised regions of phase space. A similar study has already been performed for a four-body $D$ decay to $\pi^{+}\pi^{-}\pi^{+}\pi^{-}$~\cite{4picisi}.

The remainder of the paper is arranged as follows. Section~\ref{sec:QC} describes the quantum-correlated $D$ mesons produced at CLEO-c. The relations used to determine $F_{+}$ as well as the strong-phase differences are also discussed here. The data set and event selection criteria are explained in section~\ref{Sec:Evt}. The $F_{+}$ and strong-phase difference calculations and results are presented in section~\ref{Sec:F+} and section~\ref{Sec:cisi}, respectively. The impact of these results on the determination of $\gamma$, illustrated with simulated experiments using the expected yield from the data set collected by the Belle II experiment, is discussed in section~\ref{Sec:sensitivity}. Section~\ref{Sec:conclusion} gives the conclusions.

\section{Quantum-correlated $\boldsymbol{D}$ mesons}
\label{sec:QC}

Decays of the vector meson $\psi$(3770) produce pairs of $D$ mesons in a $P$-wave state and hence the wave function for the decay is antisymmetric. It is possible to show that the decay rate is maximum when both the $D$ mesons decay to states of opposite $CP$ eigenvalue and zero when both have same $CP$ eigenvalue~\cite{MNayak}.
\subsection{$\boldsymbol{CP}$-even fraction $\boldsymbol{F_{+}}$}
\label{subsec:F+}

The $CP$-even fraction $F_{+}^{f}$ of a $D$ decay to final-state $f$ can be determined from samples in which both $D$ mesons are fully reconstructed; these we term "double-tagged" events. In the case that one decay is reconstructed in final state $f$ and the other to a $CP$ eigenstate $g$, the double-tagged yield in the integrated phase space can be written in terms of $F_{+}$, the $CP$ eigenvalue $\lambda_{CP}^{g}$ and the branching fractions for the states $D\to f$ and $D\to g$, $\mathcal{B}(f)$ and $\mathcal{B}(g)$ as 
\begin{equation}
M(f|g) = \mathcal{N}\mathcal{B}(f)\mathcal{B}(g) \epsilon (f|g)\left[ 1 - \lambda_{CP}^{g}(2F_{+}^{f} - 1)  \right], \label{Eq:DTY}
\end{equation}
where $\mathcal{N}$ is the number of $D^{0}\bar{D}^{0}$ pairs and $\epsilon(f|g)$ is the reconstruction efficiency. The single-tagged yield, where only one of the $D$ meson decays is reconstructed, is given by
\begin{equation}
S(g) = \mathcal{N}\mathcal{B}(g) \epsilon(g).
\end{equation}
Then $F_{+}$ can be defined as 
\begin{equation}
 F_{+}^{f} \equiv \frac{N^{+}}{N^{+}+N^{-}}. \label{Eqn:F+eq}
\end{equation}
$N^{+}$ and $N^{-}$ are $\frac{M(f|g)}{S(g)}$ for $CP$-odd and $CP$-even $g$ modes, respectively.
We can also use some multibody modes $g$ with already known $CP$-even fraction $F_{+}^{g}$, to determine $F_{+}^{f}$~\cite{4pi}:
\begin{equation}
F_{+}^{f} = \frac{N^{+}F_{+}^{g}}{N^{g} - N^{+} + 2N^{+}F_{+}^{g}}, \label{Eqn:F+PiPiPi0}
\end{equation}
where  $N^{g}$ is the ratio of double-tagged to single-tagged yields. 

The tagging mode $g$ can also be studied in bins of its own phase space, e.g. in the case of $K_{\rm S}^{0}\pi^{+}\pi^{-}$ or $K_{\rm L}^{0}\pi^{+}\pi^{-}$, such that the yield in the $i^{\mathrm{th}}$ bin of the $K^{0}_{\mathrm{S,L}}\pi^{+}\pi^{-}$ Dalitz plot is given by  

\begin{align}
\begin{split}
M_{i}(f | K_{\rm S,L}^{0}\pi^{+}\pi^{-}) = & h_{K_{\rm S,L}^{0}\pi^{+}\pi^{-}}(K_{i}^{K_{\rm  S,L}^{0}\pi^{+}\pi^{-}}+K_{-i}^{K_{\rm S,L}^{0}\pi^{+}\pi^{-}} \\
&-2c_{i}^{K_{\rm S,L}^{0}\pi^{+}\pi^{-}}\sqrt{K_{i}^{K_{\rm S,L}^{0}\pi^{+}\pi^{-}}K_{-i}^{K_{\rm S,L}^{0}\pi^{+}\pi^{-}}}(2F_{+}^{f}-1)),\label{Eqn:F+KhPiPi}
\end{split}
\end{align}
where $K_i^{K^{0}_{\rm S,L}\pi^{+}\pi^{-}}~(K^{K^{0}_{S,L}\pi^{+}\pi^{-}}_{-i})$ and $c_i^{K^{0}_{S,L}\pi^{+}\pi^{-}}$ are the fractional rate of the $D^0$ $(\bar{D}^{0})$ decays to $K^{0}_{\mathrm{S,L}}\pi^{+}\pi^{-}$  and the cosine of the average strong-phase difference, respectively, in the $i^{\mathrm{th}}$ bin~\cite{4pi}. Here, $h_{K_{\rm S,L}^{0}\pi^{+}\pi^{-}}$ is a normalization factor.

Events where both the $D$ mesons decay to the same final state $f$ also provide useful information about $F_{+}$. The double-tagged yield in that case is given by~\cite{4pi}
\begin{equation}
M(f|f) = 4\mathcal{N}\mathcal{B}(f)^{2} 
 \epsilon(f|f)F_{+}(1-F_{+}).\label{Eq:DT}
\end{equation}

\subsection{Strong-phase difference}

The sine and cosine of the amplitude weighted averages of the strong-phase difference between $D^{0}$ and $\bar{D^{0}}$, are represented as $c_{i}$ and $s_{i}$, in the decay $D\to K^{0}_{\mathrm S}\pi^{+}\pi^{-}\pi^{0}$, similar to that in $D\to K^{0}_{\mathrm S}\pi^{+}\pi^{-}$ given in section~\ref{subsec:F+}. The amplitude for $D^{0}\to f$ can be written as $A(D^{0}\rightarrow f(x))\equiv a_{x} e^{i\theta _{x}}$, where $x$ is some point in the decay phase space and $\theta_{x}$ is the strong-phase of the decay that conserves $CP$ symmetry. Similarly for $\bar{D^{0}}\to f$, the amplitude is $A(\bar{D^{0}}\rightarrow f(x))\equiv a_{\bar{x}} e^{i\theta _{\bar{x}}}$. Here, $\bar{x}$ is a point in phase space obtained by applying a $CP$ transformation to the final state system at $x$. We can define the strong-phase difference between $D^{0}$ and $\bar{D^{0}}$ as $\Delta \theta_{x} = \theta_{x} - \theta_{\bar{x}}$. So, $c_{i}$ and $s_{i}$ are defined as

\begin{equation}
c_{i} = \frac{1}{\sqrt{T_{i}\bar{T_{i}}}} \int_{x\in i} a_{x}a_{\bar{x}} \cos \Delta \theta_{x} dx
\end{equation}
and
\begin{equation}
s_{i} = \frac{1}{\sqrt{T_{i}\bar{T_{i}}}} \int_{x\in i} a_{x}a_{\bar{x}} \sin \Delta \theta_{x} dx,
\end{equation}
where $T_{i} =\int_{x\in i} a_{x}^{2} dx $ and $\bar{T_{i}} =\int_{x\in i} a_{\bar{x}}^{2} dx $.

Because of the quantum correlation, the decays of $D\to K_{\rm S}^{0}\pi^{+}\pi^{-}\pi^{0}$ recoiling against $CP$ and quasi-$CP$ eigenstates and other self-conjugate states as tag modes provide direct sensitivity to $c_{i}$ and $s_{i}$. The double-tagged yields for a $CP$ tag can be written as
\begin{equation}
M_{i}(f|CP\pm) = h_{CP}\left[ K_{i}+\bar{K_{i}} \mp 2 \sqrt{K_{i} \bar{K_{i}}} c_{i} \right], \label{Eqn:CP}
\end{equation}
where $h_{CP}$ is a normalization constant and $i$ represents a particular region of the decay phase space of $f$.
For a quasi-$CP$ tag, the $c_{i}$ sensitive term is scaled by ($2F_{+} - 1$) rather than~1. 

For the tag modes $K_{\rm S,L}^{0}\pi^{+}\pi^{-}$~\cite{KsPiPi:EPJC1, KsPiPi:EPJC2}, the double-tagged yield is
\begin{align}
\begin{split}
M_{i\pm j}(f|K_{\rm S,L}^{0}\pi^{+}\pi^{-}) = & h_{K_{\rm S,L}^{0}\pi^{+}\pi^{-}} [ K_{i} K_{\mp j}^{K_{\rm S,L}^{0}\pi^{+}\pi^{-}} + 
\bar{K_{i}} K_{\pm j}^{K_{\rm S,L}^{0}\pi^{+}\pi^{-}} \\
& - 2 \sqrt{K_{i} K_{\pm j}^{K_{\rm S,L}^{0}\pi^{+}\pi^{-}}\bar{K_{i}}K_{\mp j}^{K_{\rm S,L}^{0}\pi^{+}\pi^{-}}} ( c_{i}c_{j}^{K_{\rm S,L}^{0}\pi^{+}\pi^{-}} + s_{i}s_{j}^{K_{\rm S,L}^{0}\pi^{+}\pi^{-}})]. \label{Eqn:KSPiPicisi}
\end{split}
\end{align}
Here $j$ is a particular region of the decay phase space of $K_{\rm S,L}^{0}\pi^{+}\pi^{-}$. 
If both the $D$ meson final states are the same, then 
\begin{equation}
M_{ij}(f|f) = h_{f} \left[ K_{i}\bar{K_{j}} + \bar{K_{i}} K_{j} - 2 \sqrt{K_{i} \bar{K_{j}} \bar{K_{i}} K_{j}} (c_{i}c_{j} + s_{i}s_{j}) \right ], \label{Eqn:DTcisi}
\end{equation}
where $h_{f}$ is a normalization constant.

\section{Data set and event selection}
\label{Sec:Evt}
A data sample consisting of $D\bar{D}$ pairs coming from the $\psi(3770)$ resonance collected by the CLEO-c detector at the CESR $e^{+}e^{-}$ collider  is used in this analysis. This corresponds to an integrated luminosity of 0.82~fb$^{-1}$. A detailed description of the CLEO-c detector is given in refs.~\cite{CLEO1,CLEO2,CLEO3,CLEO4}. Monte Carlo (MC) simulations of signal events are used to estimate selection efficiencies. Generic samples of $D\bar{D}$ MC events having twenty times the integrated luminosity of the data set are used to determine the background contributions. The {\tt EvtGen}~\cite{Evtgen} package is used to generate the decays and the detector response is modelled with {\tt Geant}~\cite{Geant}. The final-state radiation effects associated with charged particles are simulated with {\tt PHOTOS}~\cite{PHOTOS}.

 One of the $D$ mesons is reconstructed in the final state of interest, $K_{\rm S}^{0}\pi^{+}\pi^{-}\pi^{0}$, and the other to one of the different tag states given in table~\ref{Table:Tags}. All tracks and showers associated with both the $D$ mesons are reconstructed; the selection criteria for the tag modes are identical to those presented in ref.~\cite{MNayak}. 
 
\begin{table} [t] 
\centering  
 \begin{tabular} {|c| c|  }
\hline 
Type & Modes \\[0.5ex]
\hline
\hline
 $CP$-even & $K^{+}K^{-}$, $\pi^{+}\pi^{-}$, $K_{\rm S}^{0}\pi^{0}\pi^{0}$, $K_{\rm L}^{0}\omega$, $K_{\rm L}^{0}\pi^{0}$ \\[0.5ex]
 $CP$-odd & $K_{\rm S}^{0}\pi^{0}$, $K_{\rm S}^{0}\eta$, $K_{\rm S}^{0}\eta'$ \\[0.5 ex]
 Mixed $CP$ & $\pi^{+}\pi^{-}\pi^{0}$, $K_{\rm S}^{0}\pi^{+}\pi^{-}$, $K_{\rm L}^{0}\pi^{+}\pi^{-}$ \\[0.5 ex]
 Flavour & $K^{\pm}e^{\mp}\nu_{\rm e}$ \\[0.5ex]
\hline
\end{tabular}
\caption{Different tag modes used in the analysis.}\label{Table:Tags}
\end{table}

\begin{figure}[t]
\centering
\includegraphics[width=7cm, height =6cm]{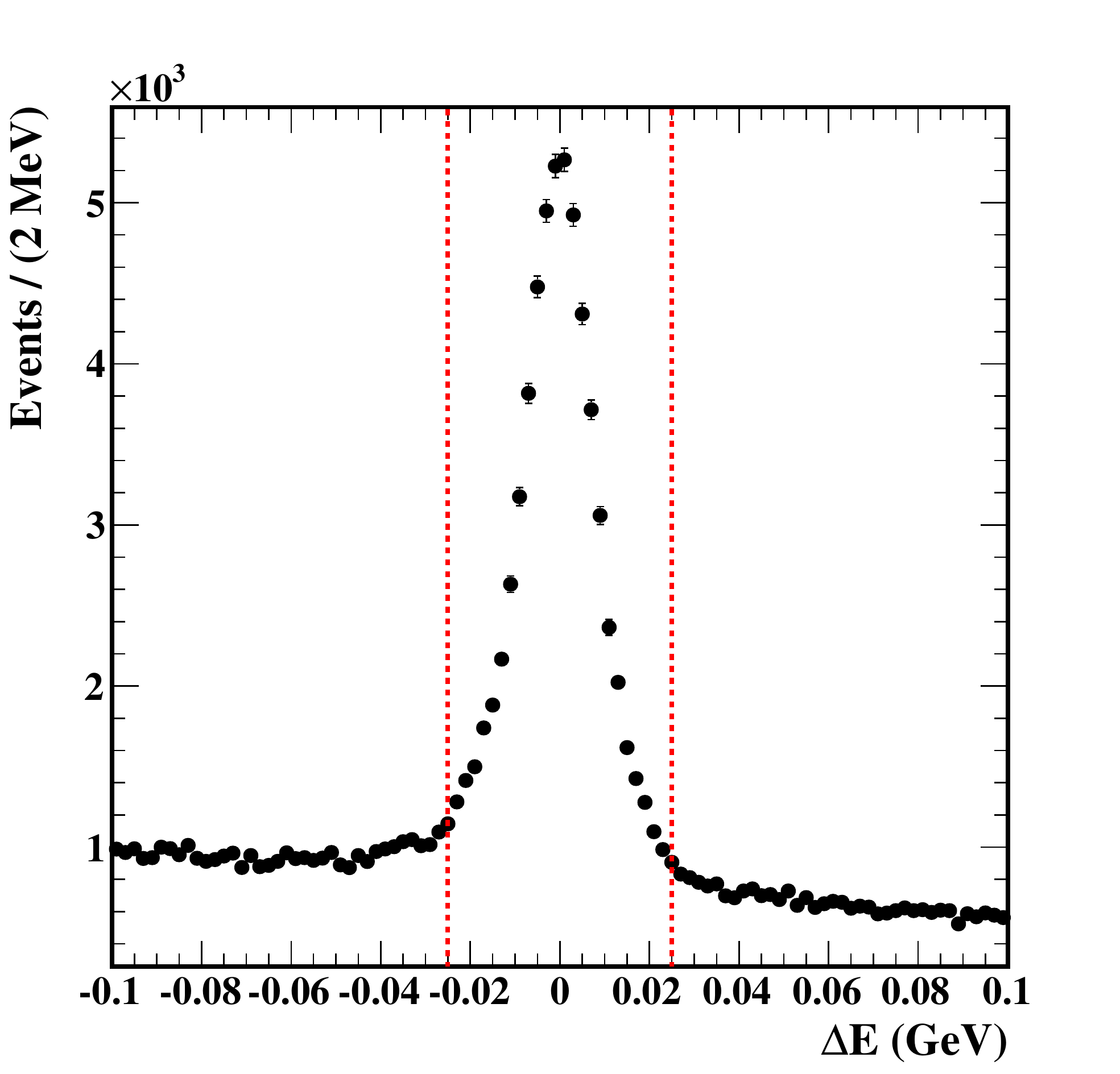}
\caption{$\Delta E$ distribution for $K_{\rm S}^{0}\pi^{+}\pi^{-}\pi^{0}$ single-tagged candidates. The vertical dotted lines indicate the signal region.}\label{Fig:STde}
\end{figure}

\begin{figure}[t]
\centering
\begin{tabular}{ccc}
\includegraphics[width=4.5cm, height =4cm]{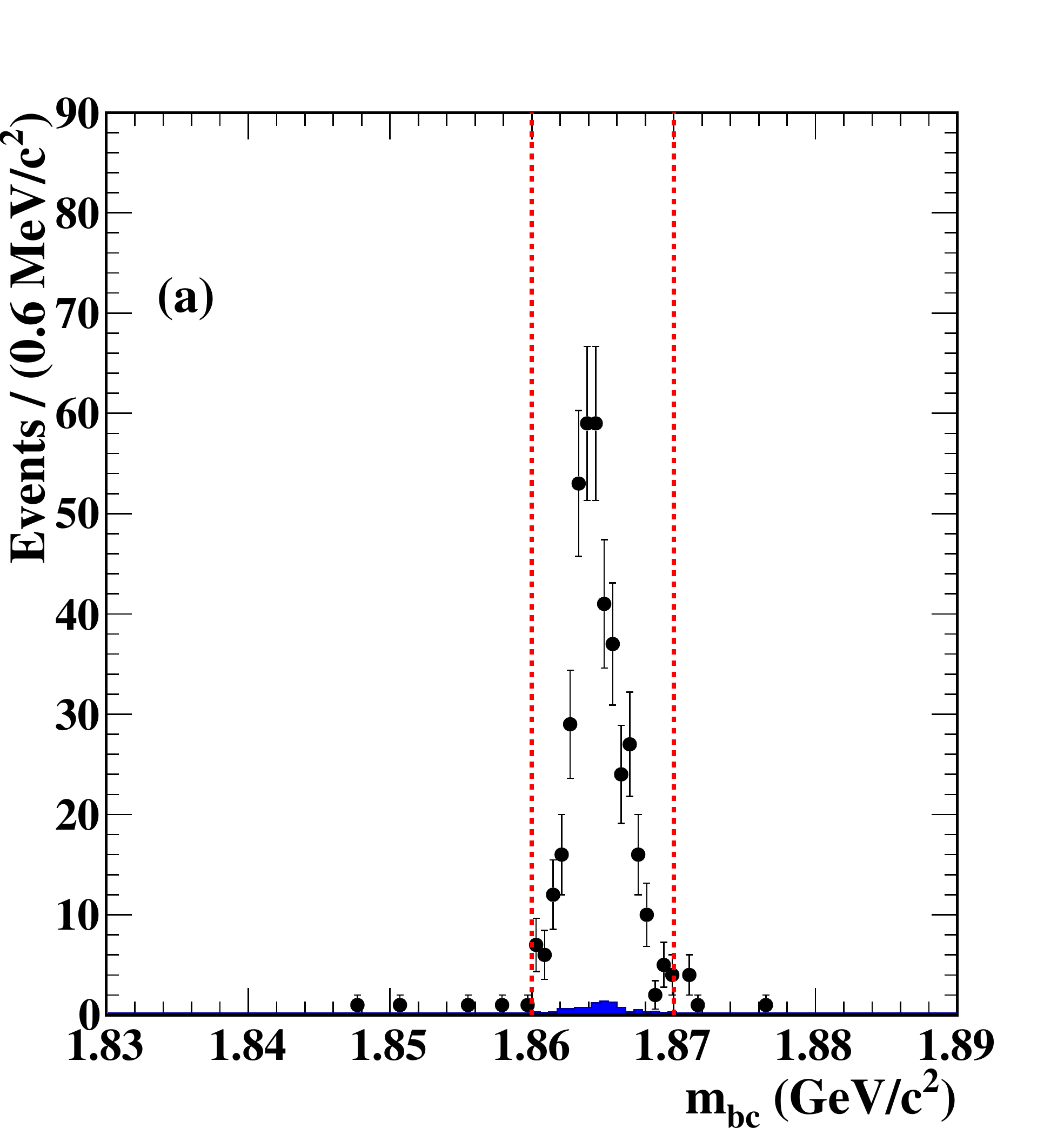}&
\includegraphics[width=4.5cm, height =4cm]{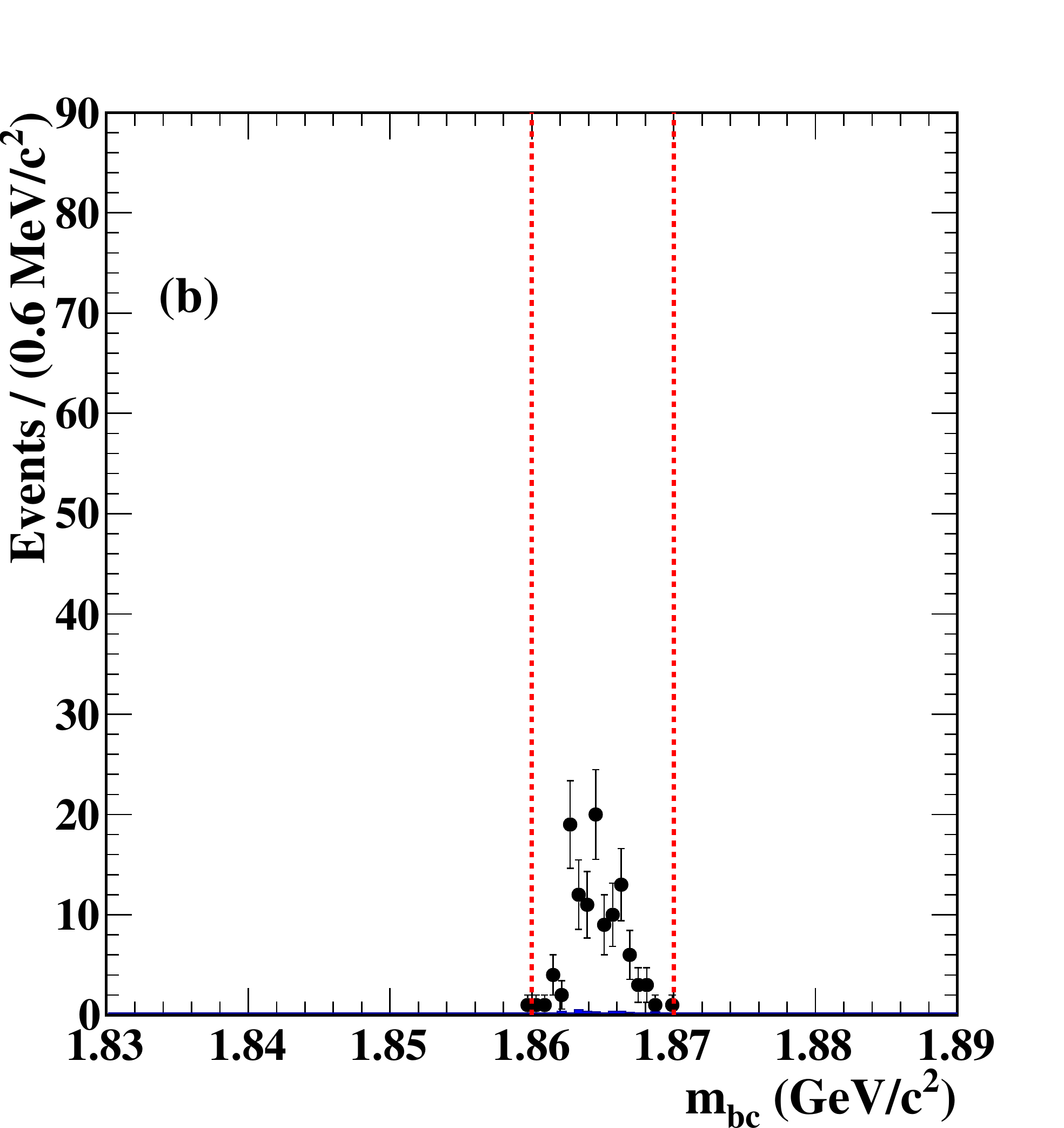} &
\includegraphics[width=4.5cm, height =4cm]{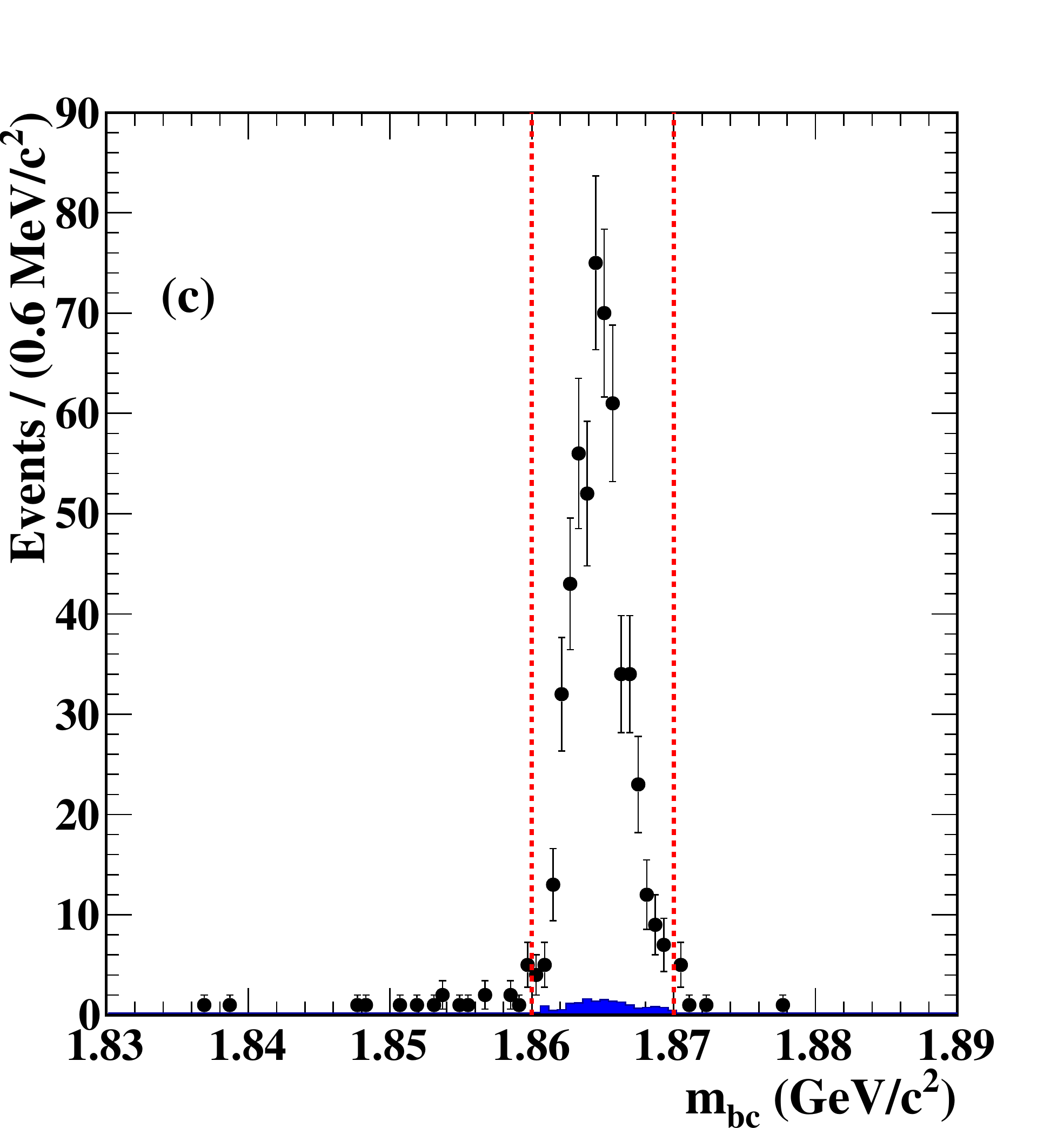}\\
\end{tabular}
\caption{$m_{bc}$ distributions for $D\to K_{\rm S}^{0}\pi^{+}\pi^{-}\pi^{0}$ decays  tagged by (a) $CP$-even states, (b) $CP$-odd states both not involving a $K_{\rm L}^{0}$ meson and (c) $K_{\rm S}^{0}\pi^{+}\pi^{-}$. The shaded histogram shows the estimated peaking background and the vertical dotted lines indicate the signal region.}\label{Fig:Mbc}
\end{figure}

\begin{figure}[t]
\centering
\includegraphics[width=7cm, height =6cm]{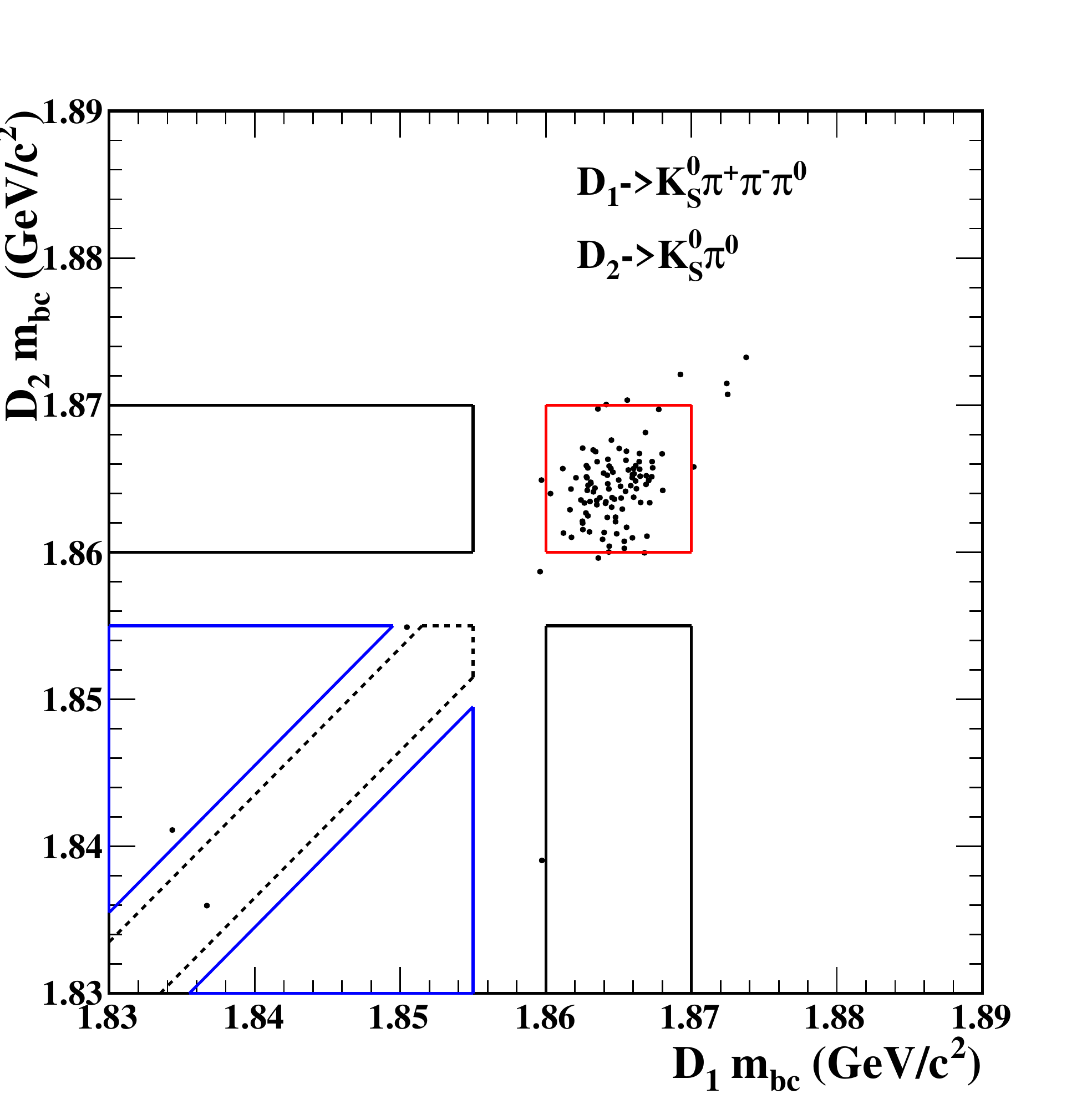}
\caption{Two dimensional $m_{bc}$ plane distribution for $K_{\rm S}^{0}\pi^{+}\pi^{-}\pi^{0}$ tagged with $K_{\rm S}^{0}\pi^{0}$ decays. The red square box indicates the signal region and the remaining boxes show the various sideband regions that are used to determine the combinatorial background contribution.}\label{Fig:Mbcplane}
\end{figure}

Hadronic modes not involving a $K_{\rm L}^{0}$ or $\nu$ are fully reconstructed using the kinematic variables beam-constrained mass ($m_{bc}$) and the beam-energy difference ($\Delta E$), which are defined as
\begin{align}
m_{bc} &= c^{-2}\sqrt{E_{\rm beam}^{2}-|\vec{P_{D}}|^2c^{2}} \\
\Delta E &= E_{D} - E_{\rm beam},
\end{align}
where $E_{\rm beam}$ is the beam energy and $\vec{P_{D}}$ and $E_{D}$ are the summed momenta and energy of the $D$ daughter particles, respectively. A kinematic fit is performed to constrain the final state particles to the $D$ meson invariant mass. This fit improves the momentum resolution of the $D$ daughter particles. For a correctly reconstructed $D$ meson, $m_{bc}$ and $\Delta E$ peak at the nominal $D$ mass~\cite{PDG} and zero, respectively. No peaking structure is observed for combinatorial backgrounds. The double-tagged yield is calculated by counting the events in the signal and sideband regions of $m_{bc}$ with different selection criteria on $\Delta E$ for various modes as in ref.~\cite{MNayak}. For the mode $K_{\rm S}^{0}\pi^{+}\pi^{-}\pi^{0}$, not previously analysed, we impose $-0.025~\rm GeV<\Delta E<0.025~\rm GeV$.
Figure~\ref{Fig:STde} shows the $\Delta E$ distribution for single-tagged $K_{\rm S}^{0}\pi^{+}\pi^{-}\pi^{0}$ candidates.
Double-tagged events containing two $K_{\rm S}^{0}\pi^{+}\pi^{-}\pi^{0}$ decays are also reconstructed in a similar fashion. The $m_{bc}$ distributions for $D\to K_{\rm S}^{0}\pi^{+}\pi^{-}\pi^{0}$ decays tagged with $CP$ eigenstates, not involving a $K_{\rm L}^0$, and $K_{\rm S}^{0}\pi^{+}\pi^{-}$ are shown in figure~\ref{Fig:Mbc}. An example of the two-dimensional $m_{bc}$ plane distribution for $K^{0}_{\rm S}\pi^{+}\pi^{-}\pi^{0}$ vs $K^{0}_{\rm S} \pi^{0}$  candidates is shown in figure~\ref{Fig:Mbcplane}.

\begin{figure}[t]
\centering
\begin{tabular}{cc}
\includegraphics[width=7cm, height =6cm]{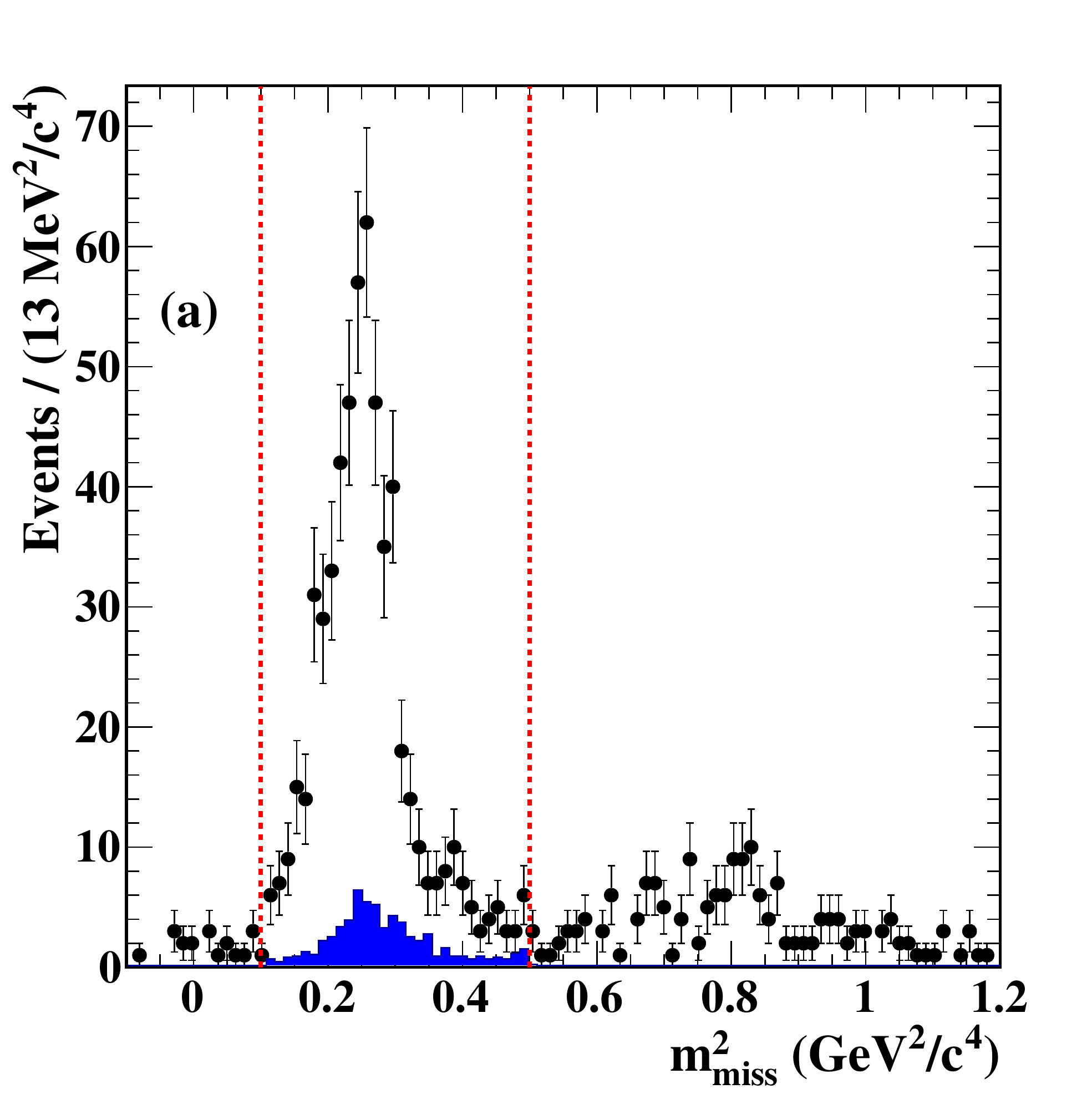}&
\includegraphics[width=7cm, height = 6cm]{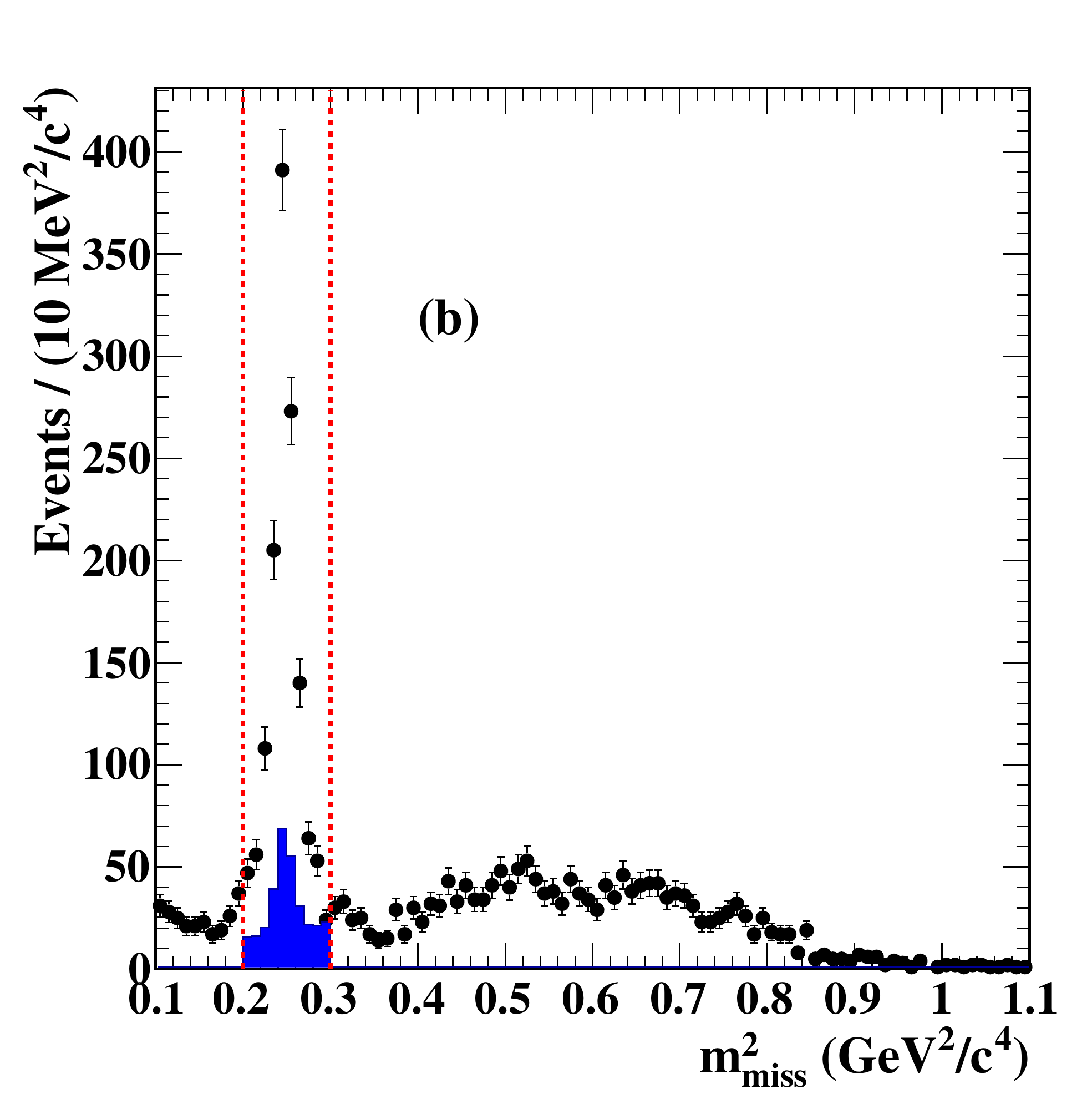} \\

\end{tabular}
\caption{$m_{\rm miss}^{2}$ distributions for $D\to K_{\rm S}^{0}\pi^{+}\pi^{-}\pi^{0}$ decays  tagged by (a) $CP$-even states involving a $K_{\rm L}^{0}$ meson and (b) $K_{\rm L}^{0}\pi^{+}\pi^{-}$. The shaded histogram shows the estimated peaking background and the vertical dotted lines indicate the signal region.}\label{Fig:MissMass}
\end{figure}

\begin{figure}[t]
\centering
\includegraphics[width=7cm, height =6cm]{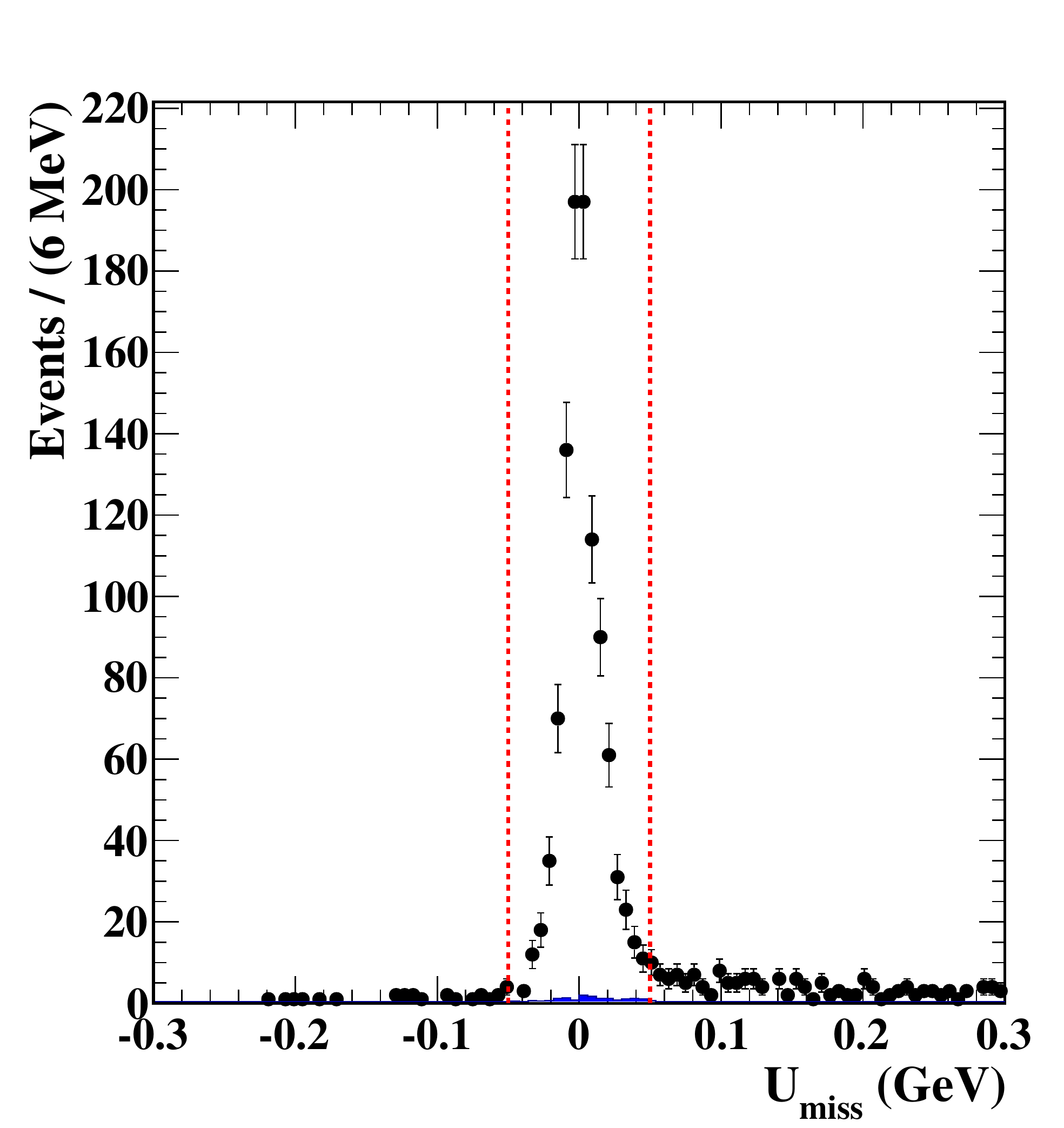}
\caption{$U_{\rm miss}$ distribution for $K^{\pm}e^{\mp}\nu_{e}$ tag. The shaded histogram shows the estimated peaking background and the vertical dotted lines indicate the signal region.}\label{Fig:UMiss}
\end{figure}

\begin{figure}[t]
\centering
\begin{tabular}{cc}
\includegraphics[width=7cm, height =6cm]{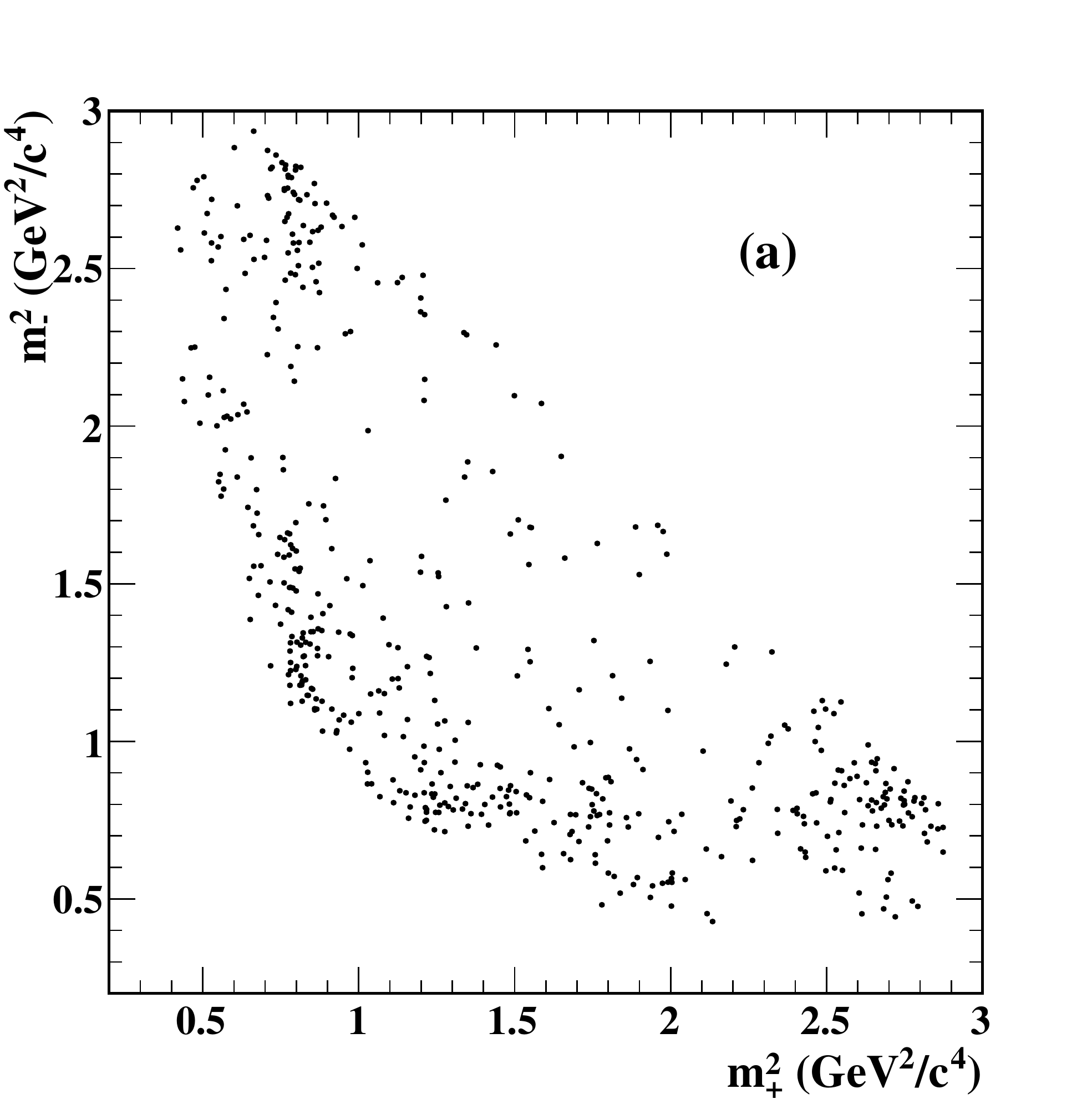}&
\includegraphics[width=7cm, height =6cm]{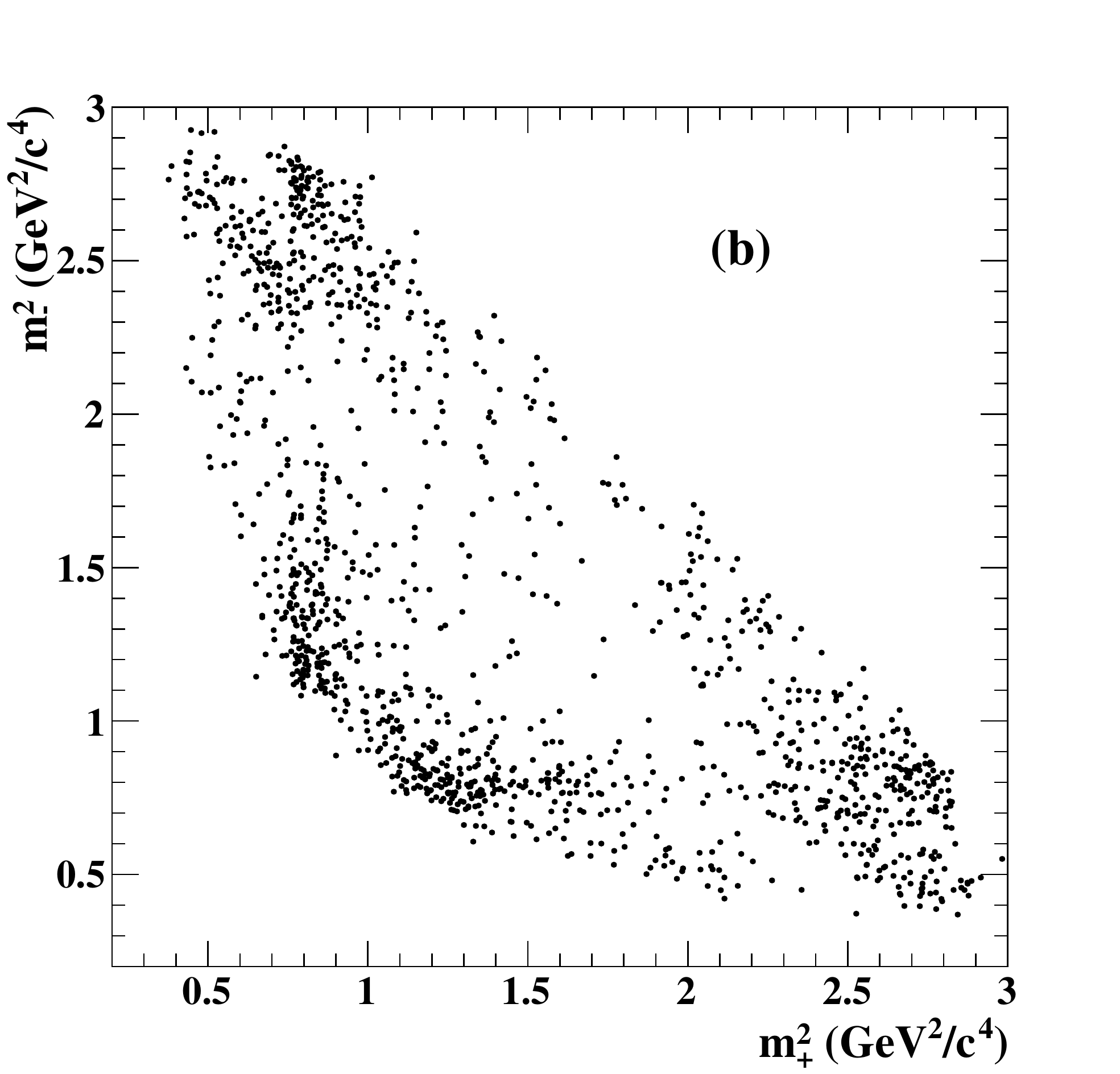}\\
\end{tabular}
\caption{Dalitz plot distributions for the tags (a) $K_{\rm S}^{0}\pi^{+}\pi^{-}$ and (b) $K_{\rm L}^{0}\pi^{+}\pi^{-}$ against $D\to K_{\rm S}^{0}\pi^{+}\pi^{-}\pi^{0}$ decays.  The axis labels $m_{\pm}^{2}$ represents the invariant mass squares of $K_{\rm S,L}^{0}\pi^{\pm}$ pairs.}\label{Fig:Dalitz}
\end{figure}

Modes with a $K_{\rm L}^{0}$ meson in the final state cannot be reconstructed fully because the $K_{\rm L}^{0}$ escapes the detector before leaving any useful signature. Hence a missing-mass squared technique~\cite{MissMass} is used to reconstruct those events. The $m_{\rm miss}^{2}$ is calculated as
\begin{equation}
m_{\rm miss}^{2} = E_{\rm miss}^{2}c^{-4} - P_{\rm miss}^{2}c^{-2},
\end{equation}
where $E_{\rm miss}$ is the missing energy and $P_{\rm miss}$ is the magnitude of the missing three-momentum in the event. For a correctly reconstructed event, $m_{\rm miss}^{2}$ peaks near the square of the  $K_{\rm L}^{0}$ mass~\cite{PDG}. The double-tagged yields are estimated from the signal and sideband regions of $m_{\rm miss}^{2}$ distribution. Similarly, semileptonic decays involving a neutrino are reconstructed by considering the quantity
\begin{equation}
U_{\rm miss} = E_{\rm miss} - c P_{\rm miss}\;,
\end{equation}
which peaks near zero for a correctly selected event. The yield in this category is estimated by looking at the signal and sideband regions of $U_{\rm miss}$ distribution. The $m_{\rm miss}^{2}$ distributions for $D\to K_{\rm S}^{0}\pi^{+}\pi^{-}\pi^{0}$ decays  tagged by $CP$-even states involving a $K_{\rm L}^{0}$ meson and $K_{\rm L}^{0}\pi^{+}\pi^{-}$ along with the $U_{\rm miss}$ distribution for $K^{\pm}e^{\mp}\nu_{e}$ tag are shown in figure~\ref{Fig:MissMass} and \ref{Fig:UMiss}, respectively. The tag-side Dalitz plot distributions for $K_{\rm S,L}^{0}\pi^{+}\pi^{-}$ are shown in figure~\ref{Fig:Dalitz}.

In events that contain more than one reconstructed pair of $D$ meson decays, the candidate with average $m_{bc}$ of both the $D$ mesons closest to the nominal mass of $D$ is chosen~\cite{PDG}. A $K_{\rm S}^{0}$ veto is applied for the final state $\pi^{+}\pi^{-}\pi^{0}$ to eliminate $K_{\rm S}^{0}(\pi^{+}\pi^{-})\pi^{0}$ background tags. As MC does not simulate quantum correlations, a correction is applied to obtain the correct amount of contamination from this source. The peaking background estimated from MC samples for modes not involving a $K_{\rm L}^{0}$ meson constitute a maximum of 2.5\% of the selected events. For $K_{\rm L}^{0}\pi^{0}$, $K_{\rm L}^{0}\omega$ and $K_{\rm L}^{0}\pi^{+}\pi^{-}$ final states, there is contamination from the corresponding $K_{\rm S}^{0}$ modes in the signal region. Again, the raw yields found in the MC are adjusted to account for quantum correlations. The mode $K^{\pm}e^{\mp}\nu$  contains no significant peaking background contributions. The background subtracted double-tagged yields for each of the modes are given in table~\ref{Table:Yield}.

\begin{table} [ht!] 
\centering
\begin{tabular} {|c |c |c|}
\hline 
Type & Mode & Yield\\[0.5ex]
\hline
\hline
$CP$-even  & $K^{+}K^{-}$ & $200.7 \pm 14.2$ \\[0.5ex]
 & $\pi^{+}\pi^{-}$ & $91.5 \pm 9.6$\\[0.5ex]
 & $K_{\rm S}^{0}\pi^{0}\pi^{0}$& $106.3 \pm 10.9$ \\[0.5ex]
 & $K_{\rm L}^{0}\pi^{0}$  & $357.3\pm20.2$\\[0.5ex]
 & $K_{\rm L}^{0}\omega$ & $162.1\pm13.7$\\[0.5ex]
\hline 
$CP$-odd & $K_{\rm S}^{0}\pi^{0}$ &$94.0 \pm 9.8$\\[0.5ex]
 &  $K_{\rm S}^{0}\eta$ & $11.6 \pm 3.7$ \\[0.5ex]
 & $K_{\rm S}^{0}\eta'$ & $7.0\pm 2.7$\\[0.5ex]
\hline
Quasi-$CP$  & $\pi^{+}\pi^{-}\pi^{0}$ &  $428.8 \pm 21.7$ \\[0.5ex]
\hline
Self-conjugate & $K_{\rm S}^{0}\pi^{+}\pi^{-}$ & $ 504.8 \pm 23.3$\\[0.5ex]
 & $K_{\rm L}^{0}\pi^{+}\pi^{-}$ & $864.1\pm46.1$\\[0.5ex]
 & $K_{\rm S}^{0}\pi^{+}\pi^{-}\pi^{0}$ & $176.4 \pm 14.8$ \\[0.5ex]
\hline
Flavour &  $K^{\pm}e^{\mp}\nu$ & $1009.8\pm32.0$\\[0.5ex]
 \hline
\end{tabular} 
\caption{Background subtracted signal yields of $D\to K_{\rm S}^{0}\pi^{+}\pi^{-}\pi^{0}$ for different tag modes.}\label{Table:Yield}
\end{table}

\begin{figure}[t]
\centering
\includegraphics[width=0.8\columnwidth]{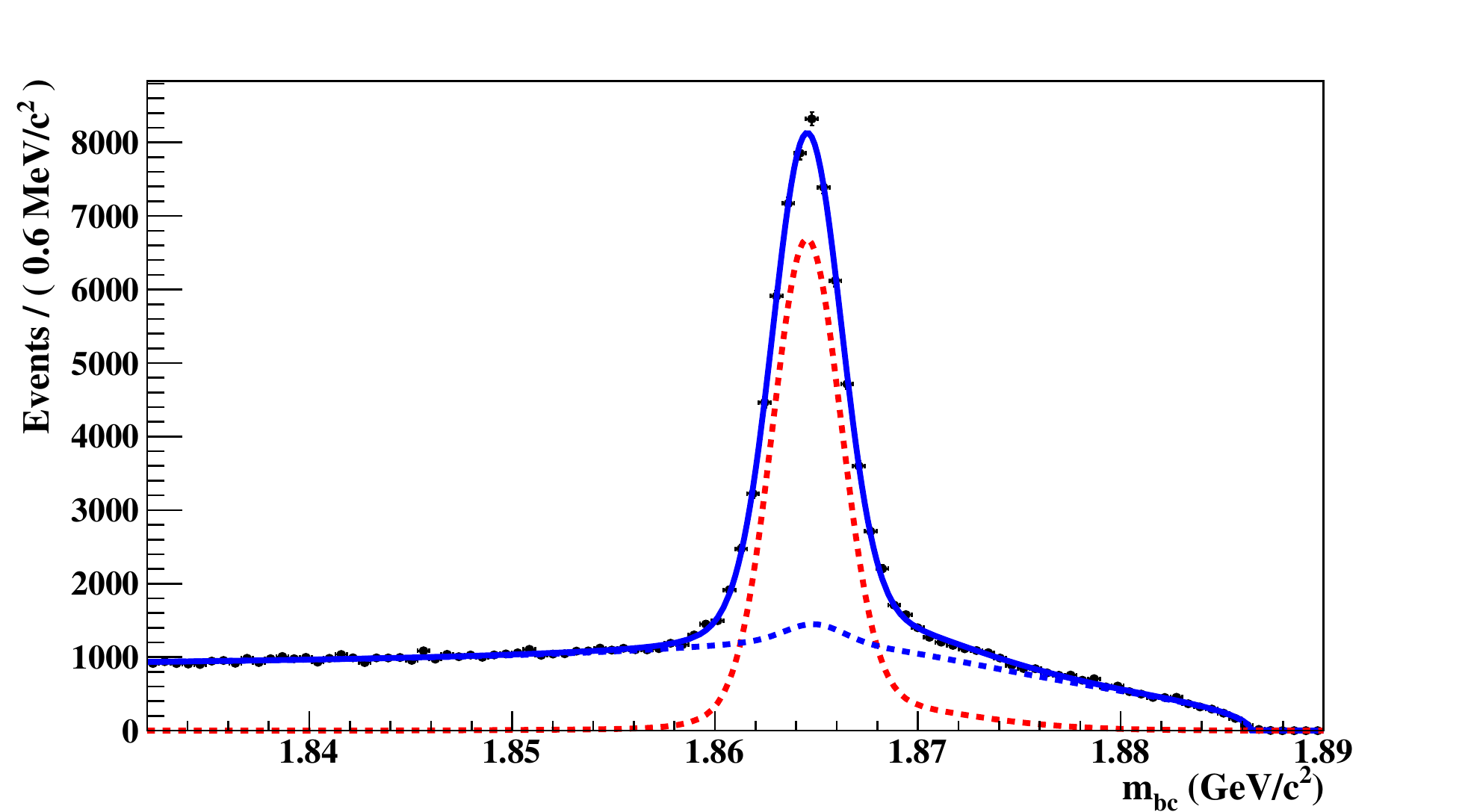}
\caption{$m_{bc}$ distribution for single-tagged $D\to K_{\rm S}^{0}\pi^{+}\pi^{-}\pi^{0}$ decays. The black points are data, the solid blue curve is the total fit and the dashed red and blue curves are signal and background fit components, respectively. }\label{Fig:datafit}
\end{figure}

The single-tagged yields for the $CP$ and quasi-$CP$ modes are taken from ref.~\cite{MNayak} as the selection criteria applied are the same. The single-tagged yield for $K_{\rm S}^{0}\pi^{+}\pi^{-}\pi^{0}$ is obtained from a fit to the $m_{bc}$ distribution and found to be 54,949~$\pm$~781. The signal component is modelled with an asymmetric Gaussian and a sum of two Gaussian probability density functions (PDF) with common mean and the background component is fitted with Argus~\cite{ARGUS}, Crystal Ball~\cite{CB} and Gaussian PDFs. The latter two PDFs in the background fit are for the small peaking component arising from $\pi^{+}\pi^{-}\pi^{+}\pi^{-}\pi^{0}$ and $K^{0}_{\mathrm{S}}K^{0}_{\mathrm{S}}\pi^{0}$. The $m_{bc}$ distribution and fit are shown in figure~\ref{Fig:datafit}.

\section{Measurement of $\boldsymbol{F_{+}}$}
\label{Sec:F+}
We perform $F_{+}$ measurements with different tag modes using the relations presented in section~\ref{subsec:F+}. The results are discussed in the following subsections.

\subsection{$\boldsymbol{CP}$ and quasi-$\boldsymbol{CP}$ tags method}
\label{Sec:CP}
\begin{figure}[t]
\centering
\begin{tabular}{cc}
\includegraphics[width=7cm]{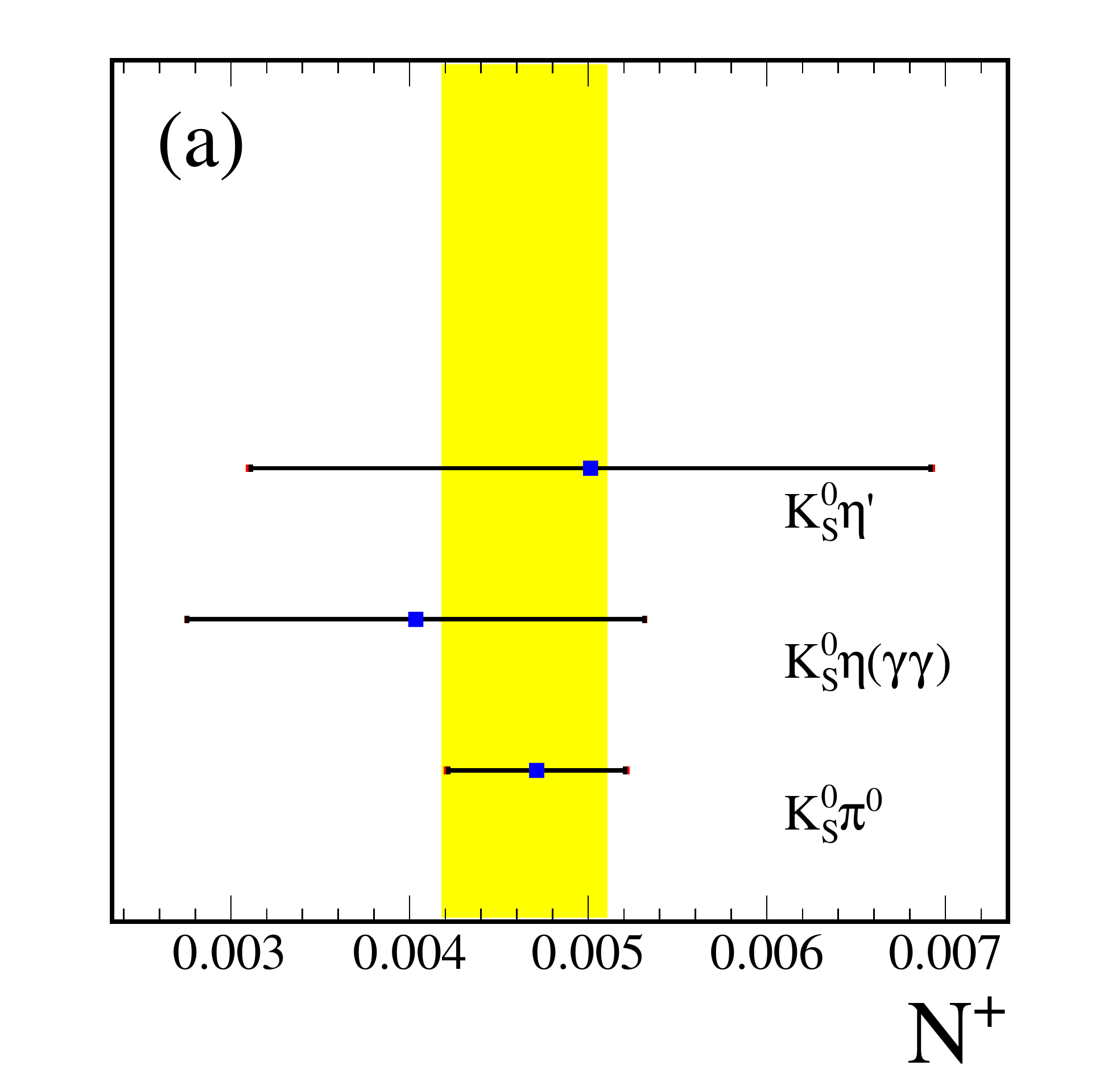}&
\includegraphics[width=7cm]{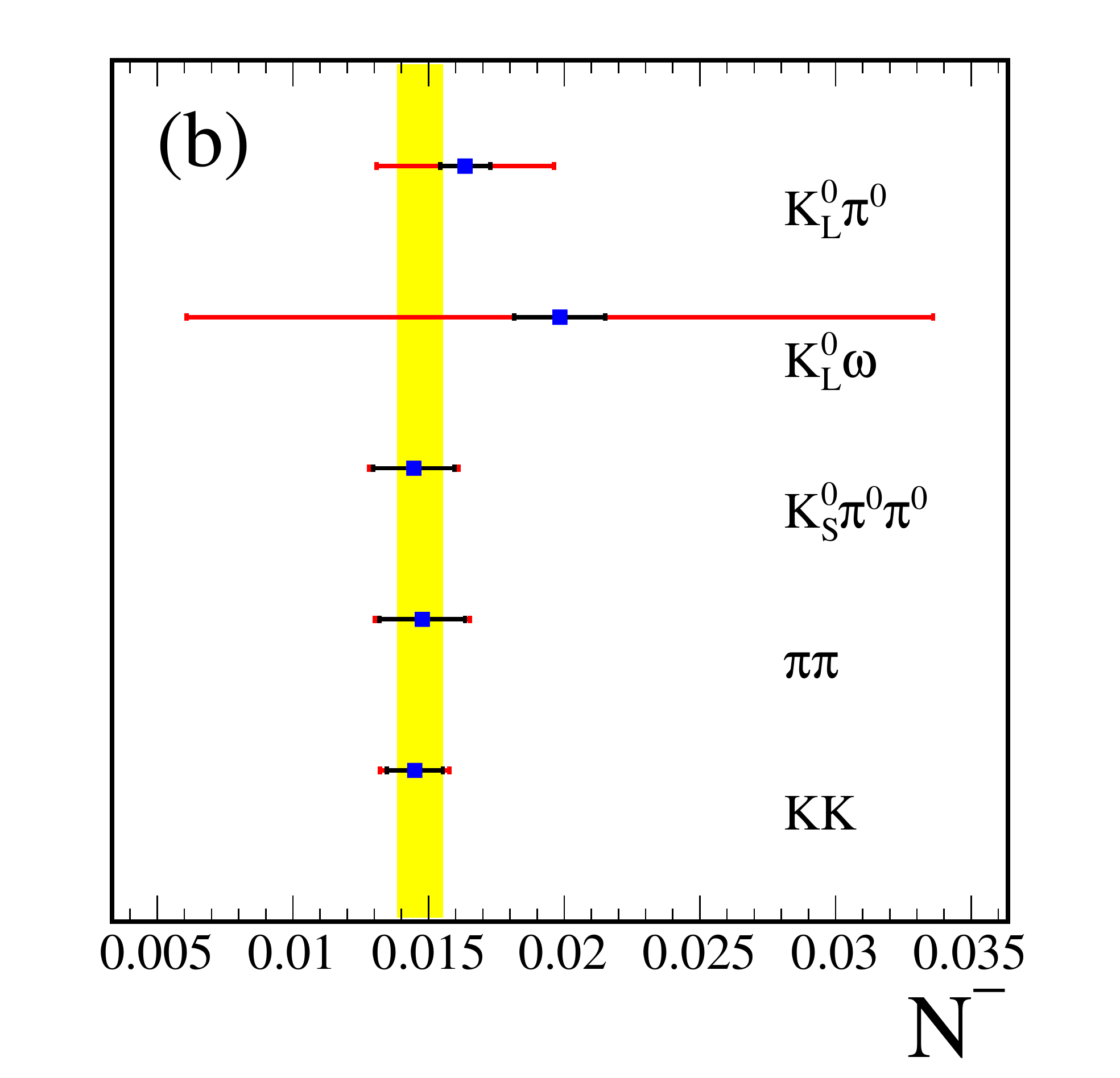}\\
\end{tabular}
\caption{ (a) $N^{+}$ values for the $CP$-odd modes and (b) $N^{-}$ values for the $CP$-even modes. The yellow region shows the average value. The horizontal black and red error bars show the statistical and the total uncertainty, respectively.}\label{Fig:N}
\end{figure}

The double-tagged yields involving a $CP$ eigenstate tag are used to obtain $N^{+}$ and $N^{-}$. 
The dependence on branching fraction and reconstruction efficiency is removed by the normalization with single-tagged yields. The possible effect of $D\bar{D}$ mixing is eliminated by applying the correction factor for single-tagged yields as $S = S_{\rm meas}/(1-\lambda_{CP}y_{D})$, where $y_{D}$ = (0.69 $\pm$ 0.06)\% is the $D$-mixing parameter~\cite{Dmix}. The $N^{+}$ and $N^{-}$ values are shown in figure~\ref{Fig:N}. It can be seen that there is consistency among the values obtained from different modes. From these results, a value of $F_{+}$~=~0.240~$\pm$~0.018~$\pm$~0.011 is calculated using eq.~\eqref{Eqn:F+eq}. The uncertainties are statistical and systematic, respectively. This value indicates that the mode $K_{\rm S}^{0}\pi^{+}\pi^{-}\pi^{0}$ is significantly $CP$-odd. The dominant systematic uncertainty comes from the determination of the single-tagged yields: in particular the fit shapes, branching fraction and reconstruction efficiency values used for $K_{\rm L}^{0}$ modes. These uncertainties are estimated in an identical manner to that described in ref.~\cite{MNayak}. 

Now, using the quasi-$CP$ tag $\pi^{+}\pi^{-}\pi^{0}$, whose $F_{+}$ value is 0.973~$\pm$~0.017~\cite{4pi}, the $CP$-even fraction for $K_{\rm S}^{0}\pi^{+}\pi^{-}\pi^{0}$ is calculated using eq.~\eqref{Eqn:F+PiPiPi0}. The result obtained with this quasi-$CP$ mode is 0.244~$\pm$~0.020~$\pm$~0.007. 

\subsection{$\boldsymbol{K_{\rm S}^{0}\pi^{+}\pi^{-}}$ and  $\boldsymbol{K_{\rm L}^{0}\pi^{+}\pi^{-}}$ tags method}

The double-tagged decays with $K_{\rm S}^{0}\pi^{+}\pi^{-}$ and $K_{\rm L}^{0}\pi^{+}\pi^{-}$ are analysed by dividing the Dalitz plot of the tag mode into eight pairs of symmetric bins as in ref.~\cite{KsPiPi} according to the amplitude model described in ref.~\cite{BaBar}. The symmetric bins are folded across the line $m_{+}^{2}=m_{-}^{2}$ to make a total of eight bins. The double-tagged yield in each of the folded bins is related to $F_{+}$ as given in eq.~\eqref{Eqn:F+KhPiPi}. Therefore $F_{+}$ can be extracted from a combined log-likelihood fit to the yields.

The background subtracted yields are determined in each of the bins for both the modes. The events in sidebands, where the tag mode is correctly reconstructed, are distributed across the Dalitz plane according to the $K_{i}$ and $\bar{K_{i}}$ values. The signal-side peaking background estimated from MC simulations are also distributed in the same manner in each of the bins. All other backgrounds are uniformly distributed across the Dalitz plane.

The reconstruction efficiency in each bin is obtained from simulated signal samples and a correction is applied to the yields to account for the variation of efficiency across the bins, which varies by typically 3\%, bin-to-bin. Table~\ref{Table:KhPiPiyield} shows the background subtracted efficiency corrected yields in each of the eight bins.

\begin{table} [ht!] 
\centering
\begin{tabular} {|c |c| c|    }
\hline 
 Bin & $K_{\rm S}^{0}\pi^{+}\pi^{-}$ & $K_{\rm L}^{0}\pi^{+}\pi^{-}$ \\[0.5ex]
\hline
\hline
 1 &  $165.8 \pm 13.5$ & $164.1 \pm 21.1 $ \\[0.5ex]
 2 &  $56.9 \pm 8.0$ & $74.7 \pm 12.9$ \\[0.5ex]
 3 & $ 46.6 \pm 7.0$& $68.7 \pm 13.7$\\ [0.5ex]
 4 &  $8.1 \pm 3.0$ & $68.6 \pm 11.1$\\ [0.5ex]
 5 & $34.0 \pm 6.0$ & $141.0 \pm 19.2$\\ [0.5ex]
 6 &  $30.4 \pm 5.8$& $86.2 \pm 14.1$ \\[0.5ex]
 7 &  $60.7 \pm 8.3$ & $131.7 \pm 16.2$ \\ [0.5ex]
 8 &  $95.3 \pm 10.1$ & $105.6 \pm 15.9$ \\[0.5ex]

\hline
\end{tabular}  
\caption{Background subtracted efficiency corrected yields of $D\to K_{\rm S}^{0}\pi^{+}\pi^{-}\pi^{0}$ decays tagged with $K_{\rm S}^{0}\pi^{+}\pi^{-}$ and $K_{\rm L}^{0}\pi^{+}\pi^{-}$ modes in bins of the tagging decay.}\label{Table:KhPiPiyield}
\end{table}

\begin{figure}[ht!]
\begin{center}
\begin{tabular}{cc}
\includegraphics[width=7.8cm]{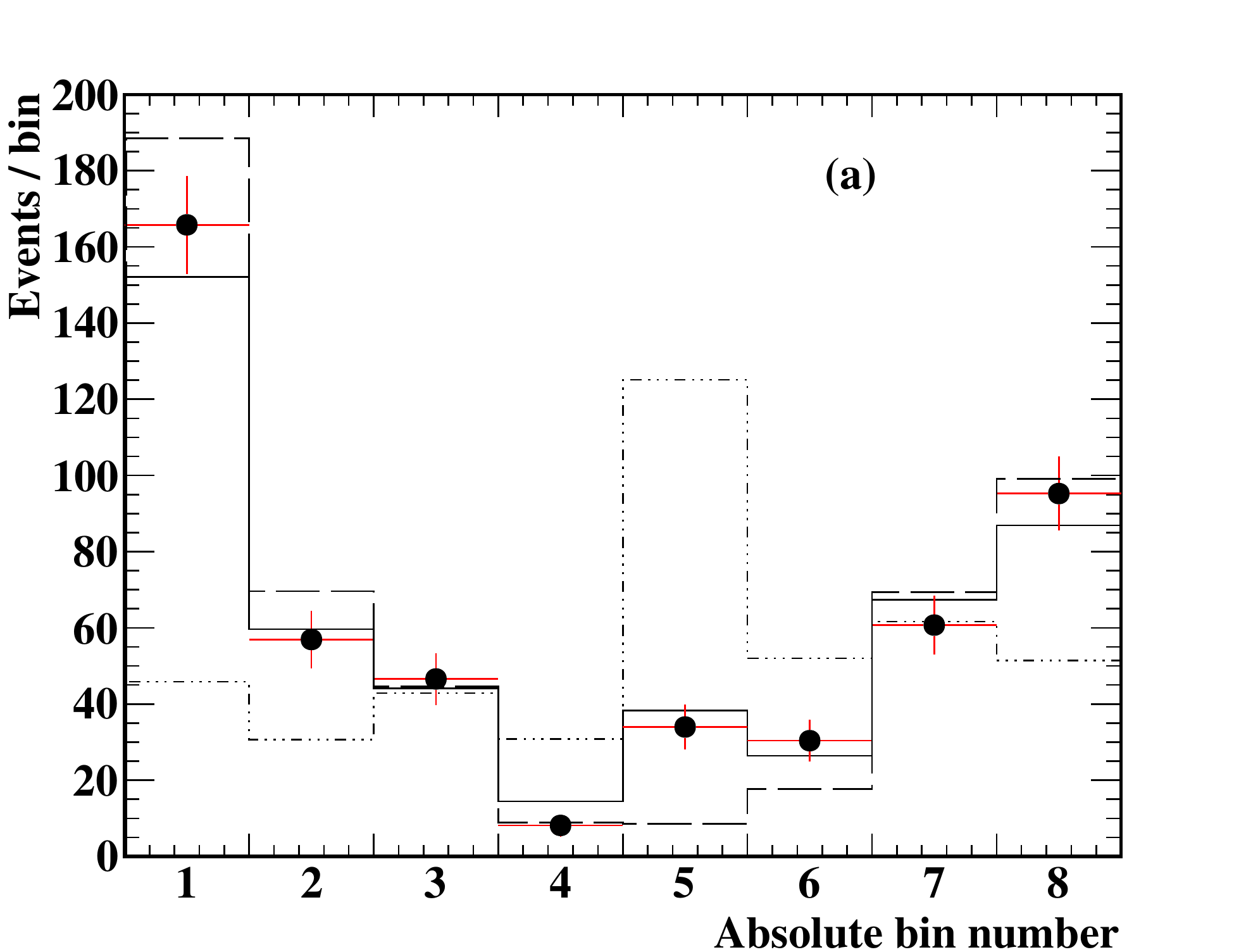}&
\includegraphics[width=7.8cm]{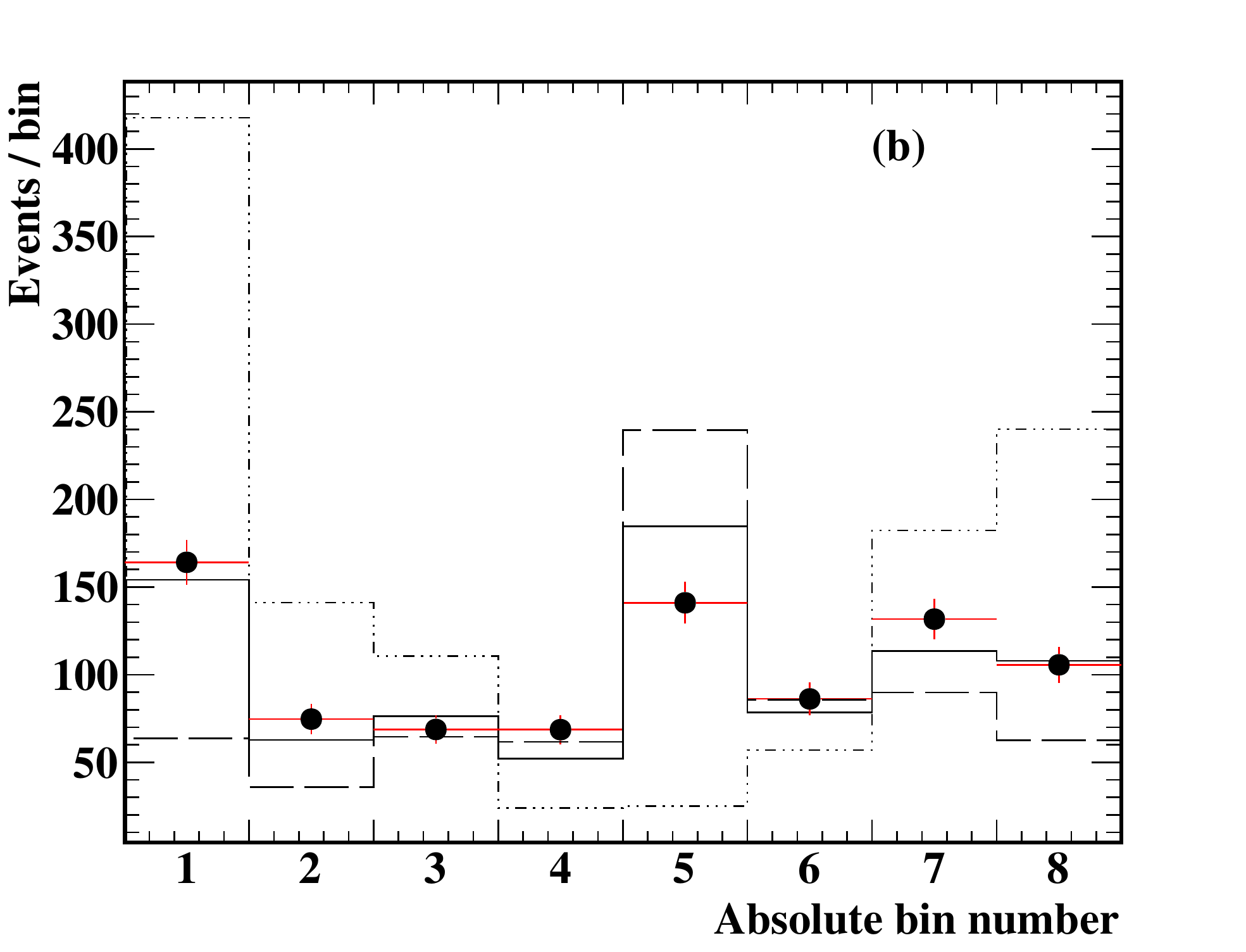}\\
\end{tabular}
\end{center}
\caption{Predicted and measured yields for (a) $K_{\rm S}^{0}\pi^{+}\pi^{-}$ and  (b) $K_{\rm L}^{0}\pi^{+}\pi^{-}$ in each bin obtained from the combined fit of both the modes. The histogram shows the predicted values from the fit, points show the measured values, the dashed line corresponds to $F_{+}$~=~0 and the dotted line shows $F_{+}$~=~1.}\label{Fig:KhPiPi}
\end{figure}

A log-likelihood fit is performed with the input yields following the form of eq.~\eqref{Eqn:F+KhPiPi} with the $CP$-even fraction and overall normalization as fit parameters. The uncertainty on the $K_{i}$, $\bar{K_{i}}$, $c_{i}$ and $c_{i}'$ input parameters are added as Gaussian constraints in the fit. The fit is performed separately for $K_{\rm S}^{0}\pi^{+}\pi^{-}$ and $K_{\rm L}^{0}\pi^{+}\pi^{-}$ and then for both the tags combined. All the fits have good quality and the results are presented in table~\ref{Table:F+KhPiPi}. The measured and predicted yields in each bin are given in figure~\ref{Fig:KhPiPi} for both tags.

\begin{table} [ht!] 
\centering
\begin{tabular} {|c |c| c|    }
\hline 
Tag & $F_{+}$& $\chi^{2}/DoF$ \\[0.5ex]
\hline
\hline
 $K_{\rm S}^{0}\pi^{+}\pi^{-}$ &  $0.194 \pm 0.040$ & 0.96 \\[0.5ex]
 $K_{\rm L}^{0}\pi^{+}\pi^{-}$ &  $0.322 \pm 0.044$ & 1.33 \\[0.5ex]
 $K_{\rm S,L}^{0}\pi^{+}\pi^{-}$ & $0.255 \pm 0.029$& 1.42\\ [0.5ex]
 
\hline
\end{tabular}  
\caption{$F_{+}$ results for the mode $K_{\rm S}^{0}\pi^{+}\pi^{-}\pi^{0}$ from the tags $K_{\rm S}^{0}\pi^{+}\pi^{-}$ and $K_{\rm L}^{0}\pi^{+}\pi^{-}$. The row $K_{\rm S,L}^{0}\pi^{+}\pi^{-}$ indicates that the combined fit includes both the samples. The fit quality metric $\chi^{2}/DoF$ is also shown, where $DoF$ stands for the number of degrees of freedom. }\label{Table:F+KhPiPi}
\end{table}

There is a two standard deviation difference between the results from each of the tags alone, however the combined result agrees with $F_{+}$ from the other tag methods. The non-uniform acceptance of the $K_{\rm S,L}^{0}\pi^{+}\pi^{-}$ Dalitz plane is studied by varying the efficiency by 3\%. The resulting change of $_{-0.008}^{+0.007}$ in $F_{+}$ is assigned as the systematic uncertainty related to this source.

\subsection{$\boldsymbol{K_{\rm S}^{0}\pi^{+}\pi^{-}\pi^{0}}$ self-tags method and combined result}

The double-tagged decays in which both the $D$ mesons decay to the same final state of $K_{\rm S}^{0}\pi^{+}\pi^{-}\pi^{0}$ can also give information about $F_{+}$ following the relation given in eq.~\eqref{Eq:DT}. The value is obtained to be 0.226~$\pm$~0.019~$\pm$~0.004. Here, the systematic uncertainty arises from the uncertainty on external input values used in the calculation such as the number of $D\bar{D}$ pairs and the branching fraction of the decay.

The value of $F_{+}$ from all these above methods are given in table~\ref{Table:F+results}. They are consistent with each other and the combined result obtained via weighted averaging is 0.238~$\pm$~0.012~$\pm$~0.003, where the correlation due to the use of $N^{+}$ for $CP$ tags  as well as the $\pi^{+}\pi^{-}\pi^{0}$ tag is taken into account.

\begin{table} [ht!] 
\centering
\begin{tabular} {|c| c|   }
\hline 
 Method & $F_{+}$\\[0.5ex]
\hline
\hline
 $CP$ tags & $0.240~\pm~0.018~\pm~0.011$\\[0.5ex]
 quasi-$CP$ tag & $0.244~\pm~0.020~\pm~0.007$\\[0.5ex]
 $K_{\rm S,L}^{0}\pi^{+}\pi^{-}$  & $0.255~\pm~0.029~^{+0.007}_{-0.008}$\\[0.5ex]
 $K_{\rm S}^{0}\pi^{+}\pi^{-}\pi^{0}$ self-tag & $0.226~\pm~0.019~\pm~0.004$\\[0.5ex]
 \hline
\end{tabular} 
\caption{$F_{+}$ results from different methods.}\label{Table:F+results} 
\end{table}

We need to consider another source of systematic uncertainty common to all methods: the non-uniform acceptance across the phase space of $D\to K_{\rm S}^{0}\pi^{+}\pi^{-}\pi^{0}$, which will bias the result with respect to the flat acceptance case. We estimate the acceptance systematic uncertainty by calculating $F_{+}$ from the $c_{i}$ strong-phase difference results given in section~\ref{Sec:cisi}, which have bin-wise efficiency corrections. The value of $F_+$ is related to $c_i$ by
\begin{equation}
F_{+} = \frac{1}{2}\left( 1 + \Sigma_{i} c_{i}\sqrt{K_{i}\bar{K_{i}}}\right).\label{Eq:F+ci}
\end{equation}
The same data are used, so any difference can be attributed to the absence of acceptance corrections in the inclusive method. The obtained result is 0.226~$\pm$~0.020. There is a one standard deviation difference between the value obtained from eq.~\eqref{Eq:F+ci} and the averaged unbinned $F_{+}$ result. The difference, 0.012, is taken as the systematic uncertainty from this source. Including this uncertainty the combined result becomes 0.238~$\pm$~0.012~$\pm$~0.012.

\section{Determination of $\boldsymbol{c_{i}}$ and $\boldsymbol{s_{i}}$}
\label{Sec:cisi}
\begin{figure}[t]
\centering
\begin{tabular}{cc}
\includegraphics[width=7cm, height =6cm]{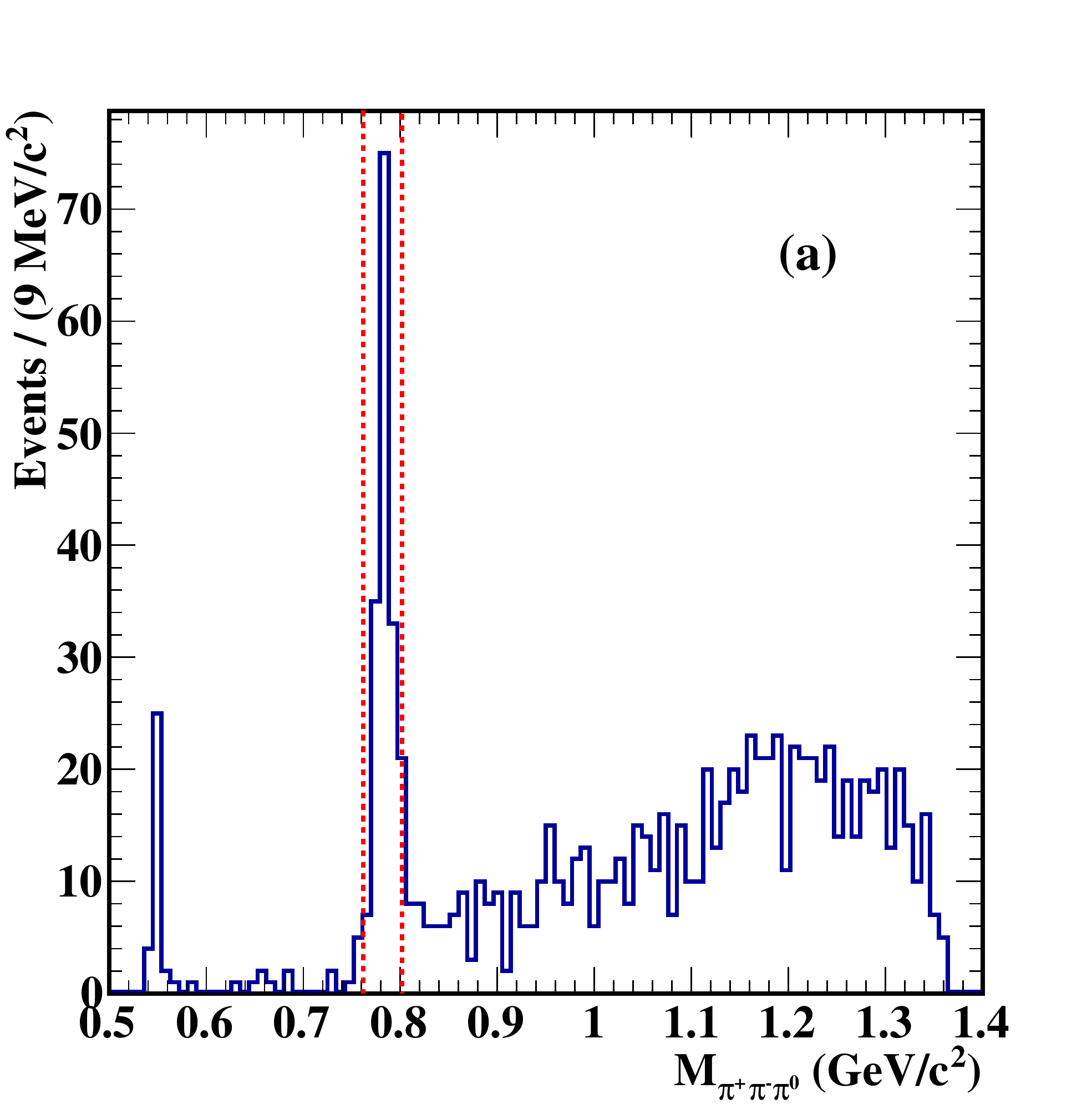}&
\includegraphics[width=7cm, height = 6cm]{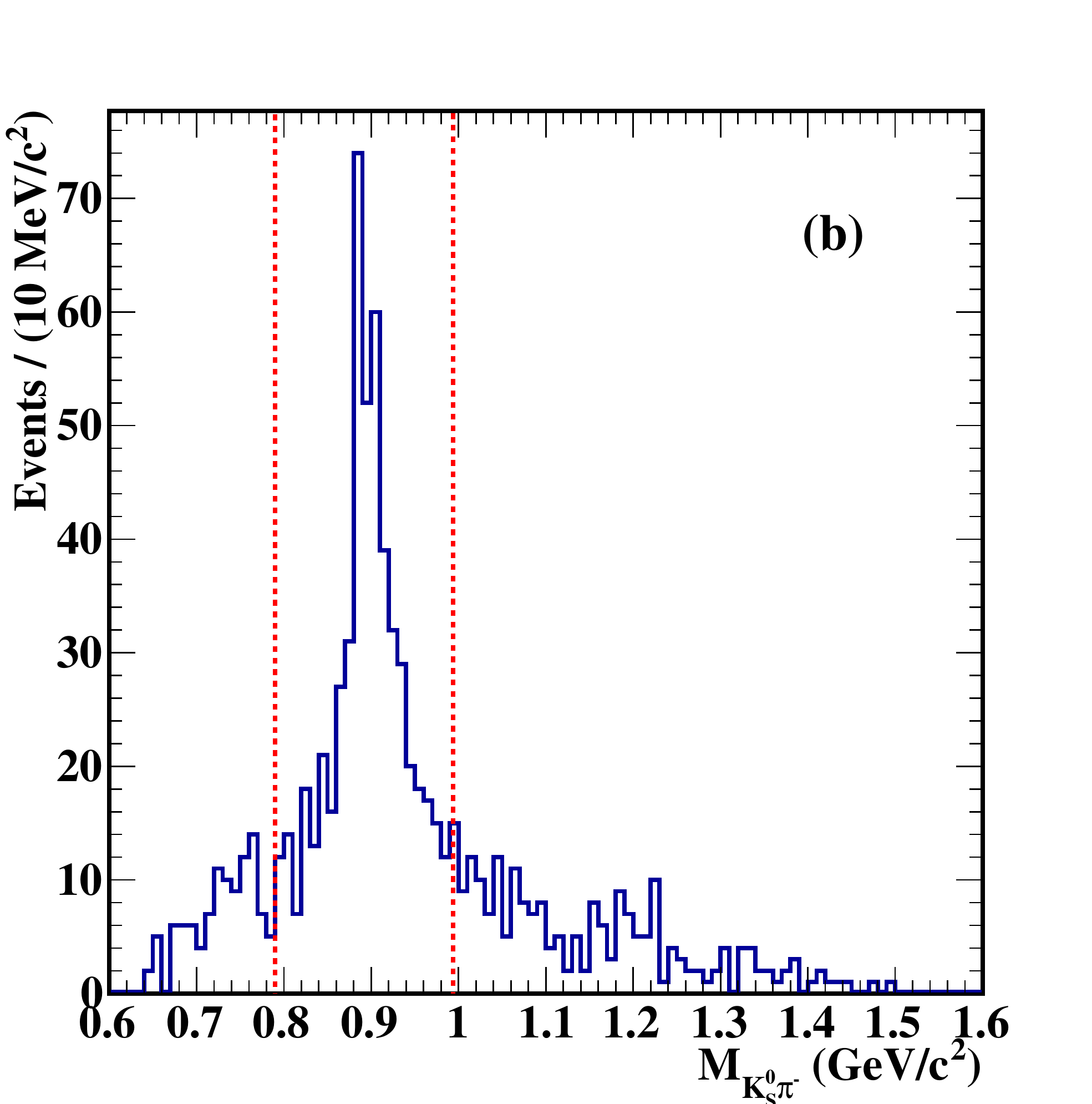} \\
\includegraphics[width=7cm, height =6cm]{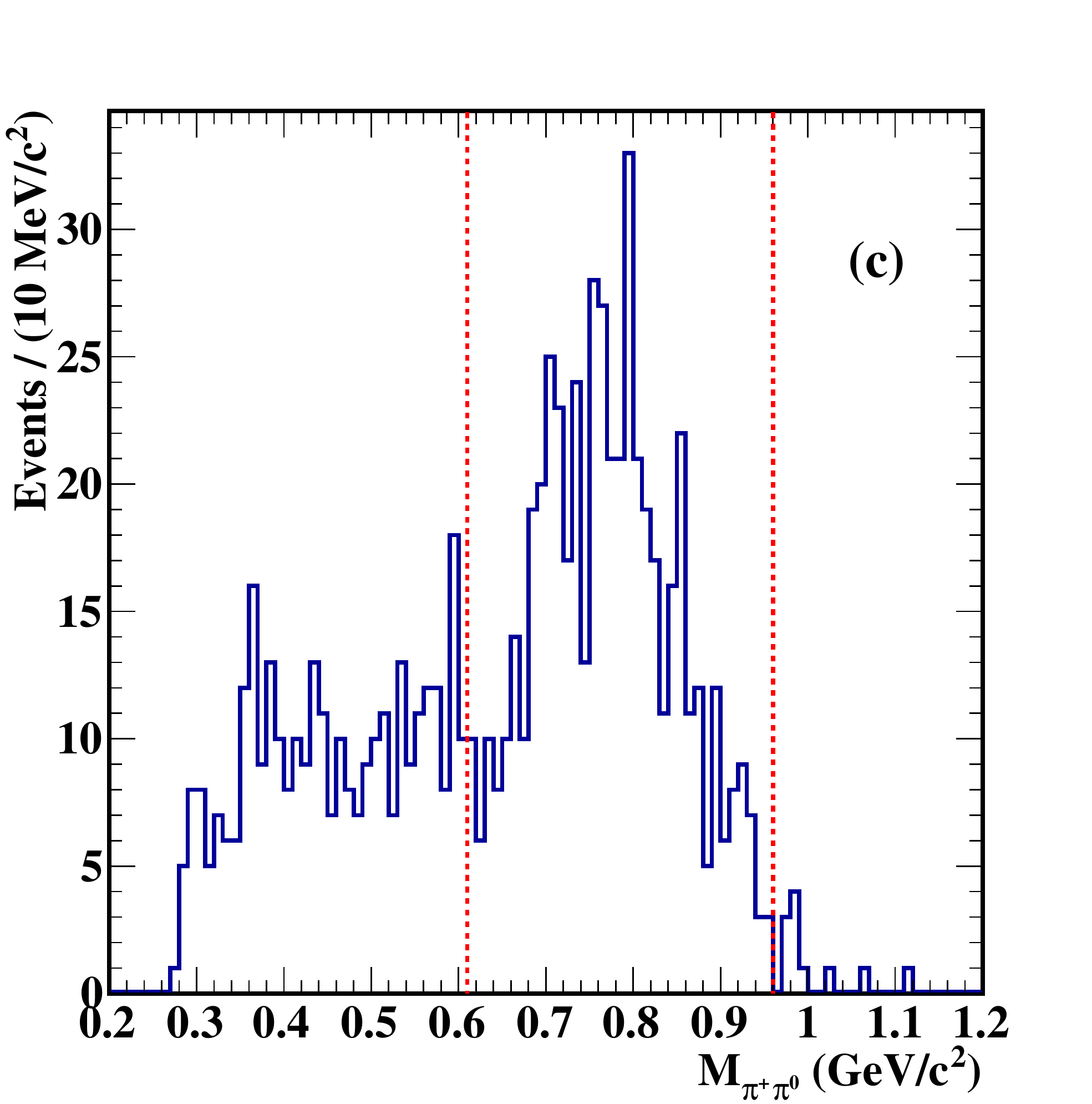}&
\includegraphics[width=7cm, height = 6cm]{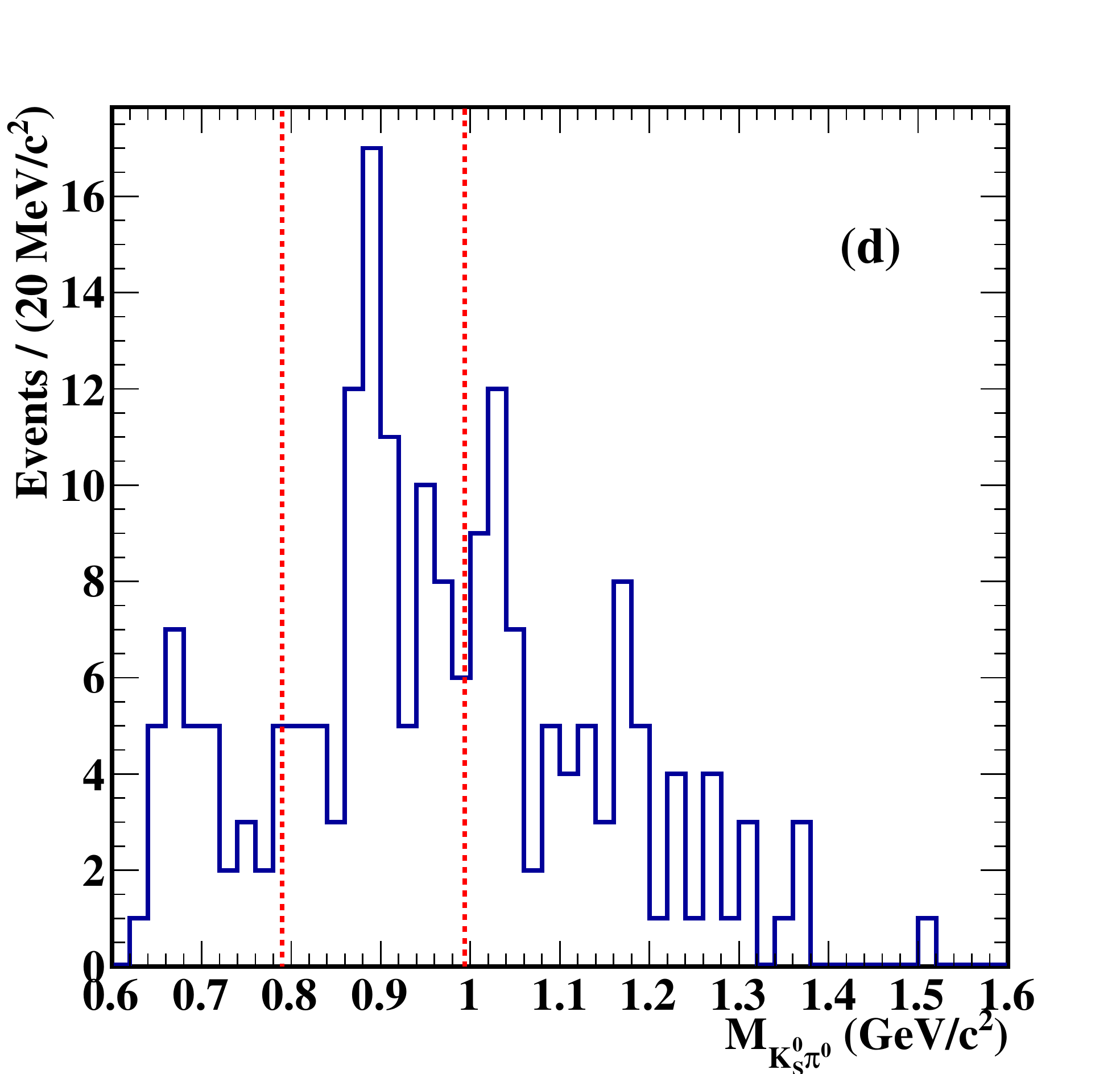} \\

\end{tabular}
\caption{Invariant mass distributions for (a) $\pi^{+}\pi^{-}\pi^{0}$, (b) $K_{\rm S}^{0}\pi^{-}$ (c) $\pi^{+}\pi^{0}$ and (d) $K_{\rm S}^{0}\pi^{0}$ of $D\to K_{\rm S}^{0}\pi^{+}\pi^{-}\pi^{0}$ decays  tagged by $K^{\pm}e^{\mp}\nu$. Candidates from the pervious bins are removed sequentially in the order given in table~\ref{Table:Bin}: (a) no events removed, (b),(c) events in bin 1 and (d) events in bins 1 to 5 removed. The vertical dotted lines indicate the selected mass windows for the $\omega$, $K^{*}$ and $\rho$ resonances, respectively.}\label{Fig:IMplots}
\end{figure}

The $c_{i}$ and $s_{i}$ values are extracted by looking at the same $D\to K_{\rm S}^{0}\pi^{+}\pi^{-}\pi^{0}$ data in bins of phase space. The decay phase space is five-dimensional, hence there is no trivial symmetry to define the bins as in the case for three-body decays. Furthermore, a proper optimization is impossible due to the lack of an amplitude model. Therefore, the bins are constructed around the most significant intermediate resonances present in the decay. A nine-bin scheme is defined around the intermediate resonances such as the $\omega$, $K^{*}$ and $\rho$. The kinematic regions of the bins are given in table~\ref{Table:Bin} along with the fractions of flavour-tagged $D^{0}$ and $\bar{D^{0}}$ decays in each of them, which are determined from  the semileptonic $D\to K^{\pm}e^{\mp}\nu_{\rm e}$ double-tagged events; the relevant kinematic distributions are shown in figure~\ref{Fig:IMplots}. The bins are exclusive and the cuts are applied sequentially in the order of the bin number. We also note that increasing the number of bins, which would result in better sensitivity to $\gamma$, led to instabilities in the fit due to the large number of null bins. MC studies led to robust results for the nine bins. We do not use $ K^{\pm}\pi^{\mp}$, $K^{\pm}\pi^{\mp}\pi^{0}$ or $K^{\pm}\pi^{\mp}\pi^{\pm}\pi^{\mp}$ as flavour tags because the corrections from the Cabibbo-favoured and doubly-Cabibbo-suppressed amplitudes cannot be calculated in the absence of an amplitude model for $D\to K_{\rm S}^{0}\pi^{+}\pi^{-}\pi^{0}$.

 \begin{table}[ht!] 
\centering
\small{
\begin{tabular} {|c| c |c| c|c|c|}
\hline 
 Bin & Bin region & $m_{\rm L}$   &$m_{\rm U}$   &$K_{i}$ & $\bar{K_{i}}$  \\[0.5ex]
 & & (GeV/c$^{2}$) & (GeV/c$^{2}$) & & \\[0.5ex]
\hline
\hline
 1 & m$_{\pi^{+}\pi^{-}\pi^{0}}$ $\approx$ m$_{\omega}$ & 0.762 & 0.802&0.2224 $\pm$ 0.0187 & 0.1768 $\pm$ 0.0168\\[0.5ex]
 2 & m$_{K_{S}^{0}\pi^{-}}$ $\approx$ m$_{K^{*-}}$ $\&$ & 0.790 & 0.994 & 0.3933 $\pm$ 0.0219 & 0.1905 $\pm$ 0.0173\\[0.5ex]
  &  m$_{\pi^{+}\pi^{0}}$ $\approx$ m$_{\rho^{+}}$ & 0.610 & 0.960 & &\\[0.5ex]
 3 & m$_{K_{S}^{0}\pi^{+}}$ $\approx$ m$_{K^{*+}}$ $\&$ & 0.790 & 0.994 &0.0886 $\pm$ 0.0128 & 0.3176 $\pm$ 0.0205\\[0.5ex]
  & m$_{\pi^{-}\pi^{0}}$ $\approx$ m$_{\rho^{-}}$ & 0.610 & 0.960& &\\[0.5ex]
 4 & m$_{K_{S}^{0}\pi^{-}}$ $\approx$ m$_{K^{*-}}$ & 0.790 & 0.994&  0.0769 $\pm$ 0.0119 & 0.0469 $\pm$ 0.0093 \\[0.5ex]
 5 & m$_{K_{S}^{0}\pi^{+}}$ $\approx$ m$_{K^{*+}}$ & 0.790 & 0.994& 0.0576 $\pm$ 0.0105 & 0.0659 $\pm$ 0.0109 \\[0.5ex]
 6 & m$_{K_{S}^{0}\pi^{0}}$ $\approx$ m$_{K^{*0}}$ & 0.790 & 0.994&  0.0605 $\pm$ 0.0107 & 0.0929 $\pm$ 0.0128\\[0.5ex]
 7 & m$_{\pi^{+}\pi^{0}}$  $\approx$ m$_{\rho^{+}}$ & 0.610 & 0.960& 0.0454 $\pm$ 0.0094 & 0.0450 $\pm$ 0.0091\\[0.5ex]
 8 & m$_{\pi^{-}\pi^{0}}$  $\approx$ m$_{\rho^{-}}$& 0.610 & 0.960& 0.0233 $\pm$ 0.0068 & 0.0195 $\pm$ 0.0061\\[0.5ex]
 9 & Remainder & - & -&0.0319 $\pm$ 0.0079 & 0.0447 $\pm$ 0.0091\\[0.5ex]
\hline
\end{tabular} 
\caption{Specifications of the nine exclusive bins of $D\to K_{\rm S}^{0}\pi^{+}\pi^{-}\pi^{0}$ phase space along with the fraction of flavour-tagged $D^{0}$ and $\bar{D^{0}}$ events in each of them.  $m_{\rm L}$ and $m_{\rm U}$ are the lower and upper limits, respectively, of the invariant masses in each region. }\label{Table:Bin}
}
\end{table}  

Due to the finite resolution of the detector, reconstructed decays may migrate to other bins in phase space. This effect is studied by looking at simulated signal events and a 9~$\times$~9 migration matrix $M$, is calculated. Each of its elements gives the ratio of the number of events reconstructed to those generated in a bin. There is a significant loss of 20\% from bin 1 due to the $\omega$ resonance having a narrow decay width. The double-tagged yields ($Y$) for each mode are corrected for the migration effects as $Y_{i} = \Sigma_{j} M_{ij}Y_{j}$, where $i$ and $j$ run from 1 to 9. The $K_{i}$ and $\bar{K_{i}}$ values given in table~\ref{Table:Bin} are also obtained after the correction applied as $K_{i} = \Sigma_{j} M_{ij}^{-1}K_{j}$. 

The background subtracted and migration corrected double-tagged yields for $CP$ tags, quasi-$CP$ tag and other self-conjugate modes are obtained in each of the bins. These inputs are used in a Poissonian log-likelihood fit with $c_{i}$ and $s_{i}$ values as fit parameters. The fit assumes that the data follow eqs.~\eqref{Eqn:CP} \textendash ~\eqref{Eqn:DTcisi}. The $CP$ and quasi-$CP$ tags provide sensitivity only to $c_{i}$ values. The tags $K_{\rm S}^{0}\pi^{+}\pi^{-}$ and $K_{\rm L}^{0}\pi^{+}\pi^{-}$ give sensitivity to both $c_{i}$ and $s_{i}$ values. The already measured strong-phase parameters for $D\to K_{\rm S,L}^0\pi^{+}\pi^{-}$  are used as inputs in the fit. As the binning scheme for the signal mode is not symmetric, it is no longer possible to exploit the symmetry of the tagging decay. Therefore, there are sixteen bins in the tag-side and nine bins in the signal-side. The sample of doubly-tagged $K_{\rm S}^{0}\pi^{+}\pi^{-}\pi^{0}$ events is also useful in providing information on $s_{i}$ values. For such events, there are nine bins each in the signal and tag side.

The uncertainties on the input strong-phase parameters of $K_{\rm S}^{0}\pi^{+}\pi^{-}$ and $K_{\rm L}^{0}\pi^{+}\pi^{-}$ are accounted as Gaussian constraints in the fit. The normalization constant in eq.~\eqref{Eqn:CP}, $h_{CP}$ is chosen for one $CP$ tag, $K^{+}K^{-}$, and all the other normalizations for events not involving $K_{\rm L}^{0}$ modes are defined as $\frac{S(tag)}{S(K^{+}K^{-})}h_{CP}$ in the fit, where $S$ represents the single-tagged yield.

The nature of the symmetry within the bins leads to certain constraints that can be imposed in the fit. Bins 1, 6 and 9 are $CP$ self-conjugate, which implies $s_{i}$ = 0. The bins 2 and 3, 4 and 5, and 7 and 8 are each $CP$-conjugate pairs, which imposes relations between their $s_{i}$ values. We have:

\begin{equation}
s_{1} = 0,~ s_{6} = 0,~ s_{9}=0; \label{Eqn:s1s6s9}
\end{equation}
\begin{equation}
s_{2}\sqrt{K_{2}\bar{K_{2}}} + s_{3}\sqrt{K_{3}\bar{K_{3}}} = 0, \label{Eqn:s2s3}
\end{equation}
\begin{equation}
s_{4}\sqrt{K_{4}\bar{K_{4}}} + s_{5}\sqrt{K_{5}\bar{K_{5}}} = 0, \label{Eqn:s4s5}
\end{equation}
\begin{equation}
s_{7}\sqrt{K_{7}\bar{K_{7}}} + s_{8}\sqrt{K_{8}\bar{K_{8}}} = 0. \label{Eqn:s7s8}
\end{equation}
In the fit, we constrain $s_{3}$, $s_{5}$ and $s_{8}$ using eqs.~\eqref{Eqn:s2s3}-\eqref{Eqn:s7s8} along with fixing $s_{1}$, $s_{6}$ and $s_{9}$ to zero. 

\subsection{Systematic uncertainties}

Several sources of systematic uncertainty are considered in the $c_{i}$ and $s_{i}$ determination. The fitter assumptions are tested using pseudo experiments. The yields are calculated for a given set of $c_{i}$ and $s_{i}$ values and they are fitted back to see the deviations in the result from the input values. The input values are given within the physically allowed region of $c_{i}^{2} + s_{i}^{2} \leq 1$. The yields in each bin are multiplied by a factor of 100, and 400 such experiments are performed. This is to avoid bias due to statistical fluctuations in certain bins where the $c_{i}$ values are unphysical. The mean of the pull distribution multiplied by the statistical uncertainty on the nominal value is taken as the systematic uncertainty due to a possible bias in the fit assumptions. The negative and positive deviations from the nominal value are summed in quadrature. The background events are fluctuated to +1$\sigma$ and $-$1$\sigma$, where $\sigma$ is the statistical uncertainty, and the fits are run to obtain $c_{i}$ and $s_{i}$ values. The difference from the nominal values are taken as the systematic uncertainty. The signal-side backgrounds are fluctuated bin by bin whereas the tag-side backgrounds are changed simultaneously for each mode owing to the correlations across the bins in signal-side. 

The limited statistics of the MC sample used to determine the migration matrix can cause variations in phase-space acceptance and biases in the results. The elements of migration matrix are smeared by +1\% and $-$1\% independently to account for this possible bias. The resulting change in $c_{i}$ and $s_{i}$ are assigned as systematic uncertainty. This 1\% deviation is large enough to take care of the effect of choosing a wrong candidate during multiple candidate selection. The single-tagged yields used in the normalization of the fit are fluctuated independently to +1$\sigma$ and $-$1$\sigma$, where $\sigma$ is the statistical uncertainty on the yield and the change in $c_{i}$ and $s_{i}$ values are taken as systematic uncertainty. 

The uncertainty on $K_{i}$ and $\bar{K_{i}}$ values is taken as a Gaussian constraint in the fit and hence no need to assign a systematic uncertainty for this. We investigate the change in migration matrix due to momentum resolution. Bin 1, which has the largest migration, hence the largest sensitivity to any data-MC discrepancy is chosen for the study. The $\pi^{+}\pi^{-}\pi^{0}$ invariant mass resolution in data and MC are found to be 5.319~$\pm$~0.064 MeV/c$^{2}$ and 4.928~$\pm$~0.003 MeV/c$^{2}$, respectively. The invariant mass in data is smeared by the quadrature difference of these two resolutions 2.019 MeV/c$^{2}$ and the migration matrix is recalculated. The effect is minimal and hence we do not assign a systematic uncertainty for this in bin~1 or in any other bins. 

The multiplicity distribution for each tag shows good agreement between data and MC. The efficiency of the selection of the best candidate in an event is $\geq$~83\% in each case. So the metric choosing the best candidate in an event does not introduce a bias. A summary of the systematic uncertainty evaluation is given in table~\ref{Table:Syst1} and \ref{Table:Syst2}. The systematic uncertainties are small compared with the statistical errors.

\begin{figure}[t]
\centering
\includegraphics[width=7cm, height =7cm]{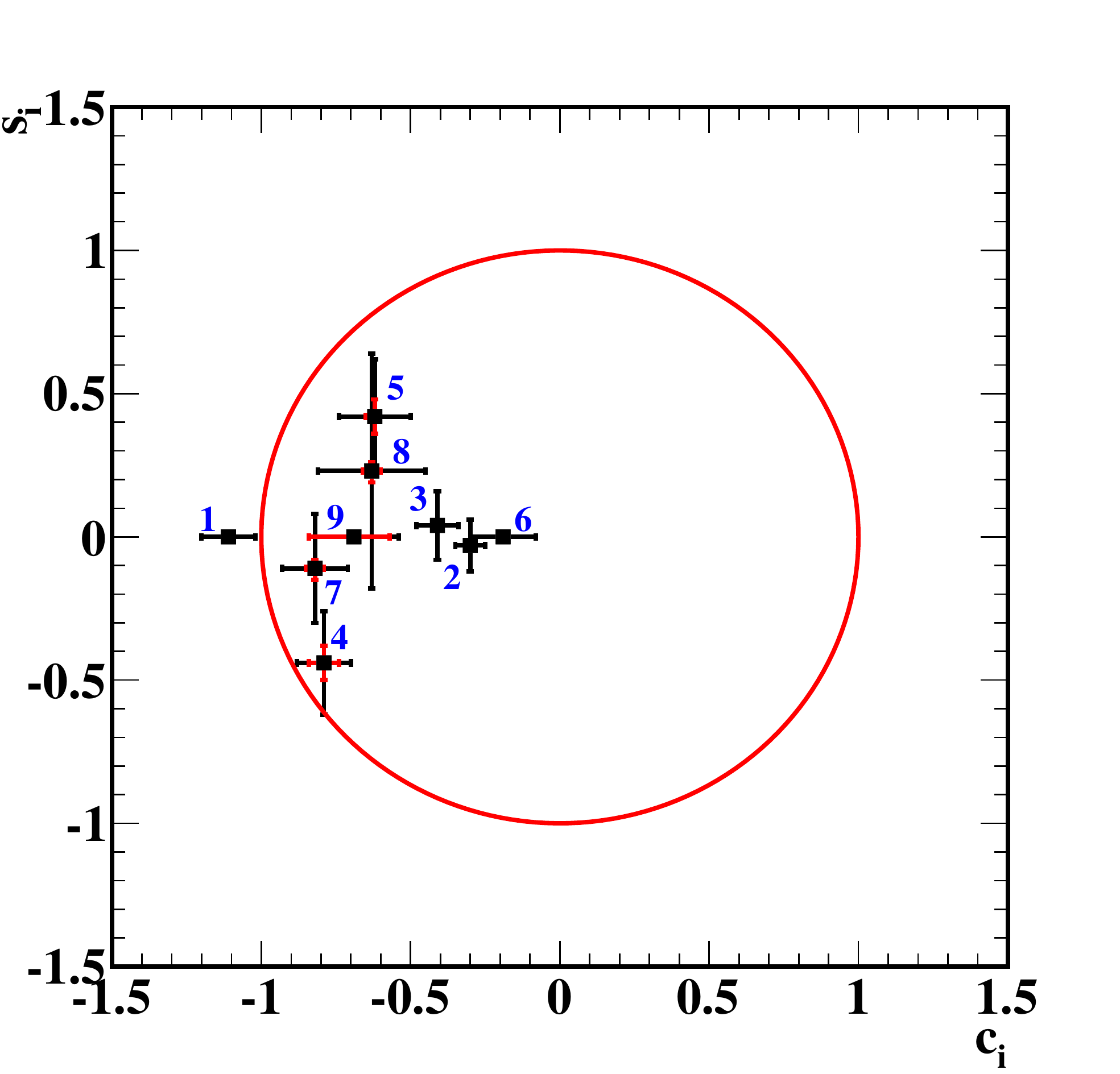}
\caption{$c_{i}$ and $s_{i}$ values in each bin. The black and red error bars represent statistical and systematic uncertainties, respectively. }\label{Fig:cisiplane}
\end{figure}

The final results of the $c_{i}$ and $s_{i}$ values are given in table~\ref{Table:finalcisi} and displayed graphically in figure~\ref{Fig:cisiplane}. The statistical and systematic correlation coefficients between $c_{i}$ and $s_{i}$ values are given in table~\ref{Table:statcorr} and table~\ref{Table:systcorr}, respectively.

\begin{table} [ht!] 

\centering
\small
\begin{tabular} {|c |c | c |c| c| c|c|c|c|c|}
\hline 
Source& $c_{1}$ & $c_{2}$& $c_{3}$& $c_{4}$& $c_{5}$& $c_{6}$& $c_{7}$& $c_{8}$& $c_{9}$ \\[0.5ex]
\hline
\hline
Fit bias & $0.003$ & $0.006$ & $0.006$ & $0.002$ & $0.002$ & $0.004$ & $0.014$ & $0.021$ & $0.046$   \\[1ex]

Peaking  & $_{-0.006}^{+0.010}$ & $_{-0.003}^{+0.004}$ &$0.005$ &$_{-0.043}^{+0.027}$ &$_{-0.011}^{+0.016}$ &$_{-0.009}^{+0.011}$ & $_{-0.015}^{+0.021}$&$_{-0.011}^{+0.013}$ &$_{-0.097}^{+0.052}$ \\[0.5ex]
background & & & & & & & & & \\[1ex]

Flat & $_{-0.011}^{+0.009}$ &$_{-0.008}^{+0.006}$ & $_{-0.010}^{+0.013}$ &$_{-0.013}^{+0.047}$ & $_{-0.021}^{+0.028}$ & $0.018$ & $_{-0.017}^{+0.023}$ & $_{-0.017}^{+0.015}$ & $_{-0.051}^{+0.131}$ \\[0.5ex]
background & & & & & & & & & \\[1ex]

Dalitz plot & $0.006$ & $ 0.002$ & $0.002$ & $_{-0.003}^{+0.002}$& $ 0.003$ & $0.002$ & $0.002$ & $0.002$ & $0.003$\\[0.5ex]
acceptance & & & & & & & & & \\[1ex]

Single-tagged & $0.001$ &$0.003$ &$0.003$ &$_{-0.002}^{+0.001}$ &$0.003$ & $0.003$ & $0.001$ & $0.002$ & $0.003$ \\[0.5ex]
yield & & & & & & & & & \\[1ex]
\hline 

Total & $_{-0.014}^{+0.015}$ & $_{-0.011}^{+0.010}$ & $_{-0.013}^{+0.015}$ & $_{-0.045}^{+0.054}$ & $_{-0.024}^{+0.032}$ & $_{-0.021}^{+0.022}$ & $_{-0.026}^{+0.034}$ & $0.029$ & $_{-0.119}^{+0.148}$ \\[1ex]
\hline
\end{tabular}
\caption{Systematic uncertainties on $c_{i}$ values.}\label{Table:Syst1} 
\end{table}

\begin{table} [ht!] 

\centering
\small
\begin{tabular} {|c |c | c |c| }
\hline 
Source& $s_{2}$ & $s_{4}$ & $s_{7}$  \\[0.5ex]
\hline
\hline

Fit bias & $0.009$ &  $0.036$ & $0.011$  \\[1ex]

Peaking  &  $0.005$   &$_{-0.032}^{+0.041}$ & $_{-0.019}^{+0.025}$ \\[0.5ex]
background & & &  \\[1ex]

Flat  &$_{-0.012}^{+0.010}$  & $_{-0.040}^{+0.031}$  & $_{-0.017}^{+0.023}$ \\[0.5ex]
background & & &  \\[1ex]

Dalitz plot &  $0.000$  & $_{-0.002}^{+0.000}$  & $0.000$ \\[0.5ex]
acceptance & & &  \\[1ex]

Single-tagged  &  $0.000$ & $_{-0.001}^{+0.000}$ &  $0.000$ \\[0.5ex]
yield& & &  \\[1ex]
\hline
Total & $_{-0.015}^{+0.014}$ & $0.063$ & $_{-0.027}^{+0.036}$ \\[1ex]

\hline

\end{tabular}
\caption{Systematic uncertainties on $s_{i}$ values.}\label{Table:Syst2} 
\end{table}

\begin{table} [ht!] 
\centering
\begin{tabular} {|c | c| c|} 
\hline 
Bin &  $c_{i}$ & $s_{i}$\\[0.5ex]
\hline
\hline
1 &         $-1.11\pm0.09_{-0.01}^{+0.02}$ & 0.00\\[0.5ex]
2 &     $-0.30\pm 0.05 \pm 0.01$ & $-0.03\pm 0.09_{-0.02}^{+0.01}$\\[0.5ex]
3 &         $ -0.41\pm0.07_{-0.01}^{+0.02}$ & $0.04\pm 0.12_{-0.02}^{+0.01~*}$\\[0.5ex]
4 &         $-0.79\pm0.09\pm 0.05$ &$-0.44\pm0.18\pm0.06$\\[0.5ex]
5 &           $-0.62\pm0.12_{-0.02}^{+0.03}$ & $0.42 \pm 0.20\pm 0.06^{~*}$ \\[0.5ex]
6 &          $-0.19\pm0.11\pm 0.02$ & 0.00\\[0.5ex]
7 &           $-0.82\pm0.11\pm 0.03$ & $-0.11\pm0.19_{-0.03}^{+0.04}$\\[0.5ex]
8 &           $-0.63\pm0.18\pm 0.03$& $0.23 \pm 0.41_{-0.03}^{+0.04~*}$ \\[0.5ex]
9 &           $-0.69\pm0.15_{-0.12}^{+0.15}$ & 0.00\\[0.5ex]
\hline
\end{tabular}
\caption{Final results for $c_{i}$ and $s_{i}$ values. The uncertainties are statistical and systematic, respectively. The $s_{i}$ results marked by * in bins 3, 5 and 8 are derived from those in other bins, according to the constraints of eqs.~\eqref{Eqn:s2s3}-\eqref{Eqn:s7s8}.} \label{Table:finalcisi}
\end{table}

\begin{table} [ht!] 
\centering
\begin{tabular} {c c c c c c c c c c c c} 
\hline 
   & $c_{2}$ & $c_{3}$ & $c_{4}$ & $c_{5}$ & $c_{6}$ & $c_{7}$ & $c_{8}$ & $c_{9}$ & $s_{2}$ & $s_{4}$ & $s_{7}$\\[0.5ex]
\hline
\hline
$c_{1}$ & 0.03 & -0.01 & -0.13 & 0.01 & 0.04 & 0.07 & -0.05 & 0.06 & 0.00 & 0.01 & 0.00 \\[0.5ex]
$c_{2}$ & & 0.08 & 0.02 & 0.06 & 0.08 & -0.02 & 0.03 & 0.06 & -0.01 & -0.01 & 0.00 \\[0.5ex]
$c_{3}$ & & & 0.03 & 0.05 & 0.01 & -0.08 & -0.03 & 0.03 & -0.03 & -0.01 & 0.01  \\[0.5ex]
$c_{4}$ & & & & -0.01 & -0.01 & -0.01 & 0.03 & -0.05 & 0.00 & -0.13 & -0.01 \\[0.5ex]
$c_{5}$ & & & & & 0.04 & 0.01 & 0.03 & 0.03 & 0.00 & -0.01 & 0.01 \\[0.5ex]
$c_{6}$ & & & & & & 0.01 & -0.03 & -0.01 & 0.00 & 0.00 & 0.00 \\[0.5ex]
$c_{7}$ & & & & & & & 0.00 & 0.00 & 0.01 & 0.00 & -0.05 \\[0.5ex]
$c_{8}$ & & & & & & & & -0.03 & -0.01 & -0.01 & 0.02 \\[0.5ex]
$c_{9}$ & & & & & & & & & 0.00 & 0.00 & 0.00 \\[0.5ex]
$s_{2}$ & & & & & & & & & & -0.03 & 0.00 \\[0.5ex]
$s_{4}$ & & & & & & & & & & & -0.02 \\[0.5ex] 
\hline
\end{tabular}
\caption{Statistical correlation coefficients between $c_{i}$ and $s_{i}$ values.} \label{Table:statcorr}
\end{table}

\begin{table} [ht!] 
\centering
\begin{tabular} {c c c c c c c c c c c c} 
\hline 
   & $c_{2}$ & $c_{3}$ & $c_{4}$ & $c_{5}$ & $c_{6}$ & $c_{7}$ & $c_{8}$ & $c_{9}$ & $s_{2}$ & $s_{4}$ & $s_{7}$\\[0.5ex]
\hline
\hline
$c_{1}$ & 0.02 & 0.02 & 0.00 & 0.02 & 0.02 & 0.00 & 0.01 & 0.01 & 0.00 & 0.00 & 0.00 \\[0.5ex]
$c_{2}$ & & 0.07 & 0.01 & 0.04 & 0.06 & 0.01 & 0.02 & 0.01 & 0.01 & 0.00 & 0.00 \\[0.5ex]
$c_{3}$ & & & 0.01 & 0.04 & 0.06 & 0.01 & 0.02 & 0.01 & 0.01 & 0.00 & 0.00  \\[0.5ex]
$c_{4}$ & & & & 0.01 & 0.01 & 0.00 & 0.00 & 0.00 & 0.00 & 0.00 & 0.00 \\[0.5ex]
$c_{5}$ & & & & & 0.05 & 0.01 & 0.01 & 0.01 & 0.01 & 0.00 & 0.00 \\[0.5ex]
$c_{6}$ & & & & & & 0.01 & 0.02 & 0.02 & 0.02 & 0.00 & 0.00 \\[0.5ex]
$c_{7}$ & & & & & & & 0.00 & 0.00 & 0.00 & 0.00 & 0.00 \\[0.5ex]
$c_{8}$ & & & & & & & & 0.00 & 0.00 & 0.00 & 0.00 \\[0.5ex]
$c_{9}$ & & & & & & & & & 0.00 & 0.00 & 0.00 \\[0.5ex]
$s_{2}$ & & & & & & & & & & 0.00 & 0.00 \\[0.5ex]
$s_{4}$ & & & & & & & & & & & 0.00 \\[0.5ex] 
\hline
\end{tabular}
\caption{Systematic correlation coefficients between $c_{i}$ and $s_{i}$ values.} \label{Table:systcorr}
\end{table}

\section{Estimation of $\boldsymbol{\gamma}$ sensitivity with $\boldsymbol{B^{\pm}\to D(K^{0}_{\rm S}\pi^{+}\pi^{-}\pi^{0})K^{\pm}}$ }
\label{Sec:sensitivity}

\begin{figure}[ht!]
\centering
\includegraphics[width=6cm, height=5cm]{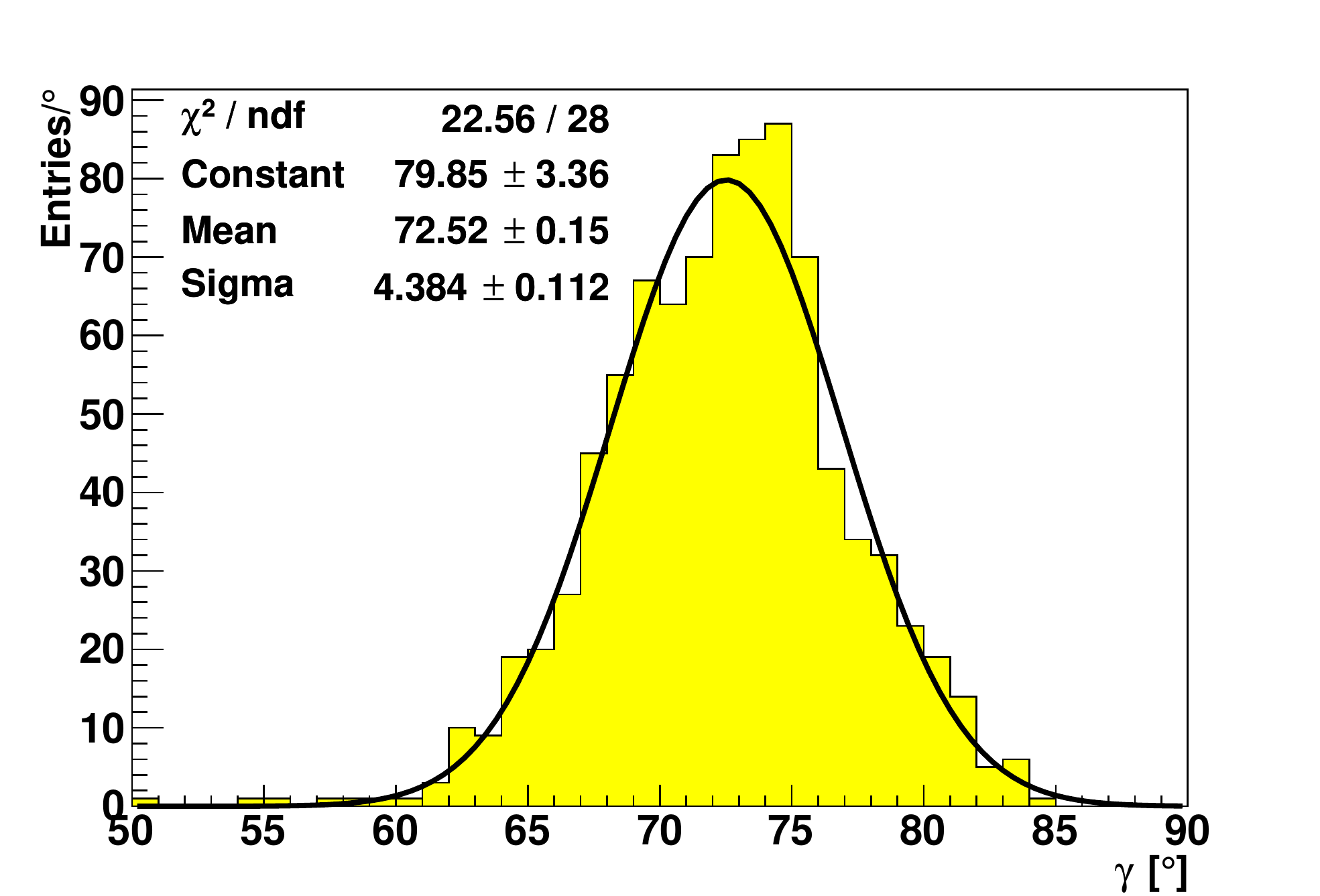}
\vspace{-0.1 in}
\caption{$\gamma$ sensitivity with 50 $\mathrm{ab}^{-1}$ Belle II sample.}\label{Fig:sensitivity}
\end{figure}

In order to estimate the impact of these results on a future $\gamma$ measurement using $B^{\pm}\to D(K^{0}_{\rm S}\pi^{+}\pi^{-}\pi^{0})K^{\pm}$ decays, we perform a simulation study based on the expected yield of this mode in the Belle data sample ($\approx$ 1~$\mathrm{ab}^{-1}$). The Belle sample of $B^{\pm}\to D(K^{0}_{\rm S}\pi^{+}\pi^{-})K^{\pm}$~\cite{Belle-GGSZ} has $\approx$ 1200 events. Assuming that increase in branching fraction for $K^{0}_{\rm S}\pi^{+}\pi^{-}\pi^{0}$ compared to $K^{0}_{\rm S}\pi^{+}\pi^{-}$ is compensated by the loss of efficiency due to a $\pi^{0}$ in the final state~\cite{Prasanth, BelleD2KsPiPi}, we expect a similar yield for $B^{\pm}\to D(K^{0}_{\rm S}\pi^{+}\pi^{-}\pi^{0})K^{\pm}$.  Then 60,000 is the yield extrapolated to the $50~\mathrm{ab}^{-1}$ data set anticipated at Belle II. The $\gamma$ sensitivity is estimated in a GGSZ~\cite{GGSZ, GGSZ2} framework. We run 1000 pseudo experiments with $c_i$, $s_i$, $K_i$, and $\bar{K_i}$ values as inputs with each experiment consisting of $\approx$ 60,000 events. The input values of $\gamma$ and the hadronic parameters $r_{B}$ and $\delta_{B}$ are taken from ref.~\cite{CKMfitter}. The estimated uncertainty on $\gamma$ is $\sigma_{\gamma}=4.4^{\circ}$ (see figure~\ref{Fig:sensitivity}). This sensitivity is very promising and only a factor two worse than that anticipated from studying $B^{\pm}\to D(K_{S}^{0}\pi^{+}\pi^{-})K^{\pm}$~\cite{b2tip} decays.

\section{Conclusions}
\label{Sec:conclusion}

Improving the knowledge of the CKM angle $\gamma$ is an important goal in flavour physics. This can be achieved by harnessing new $D$ decay modes for the measurements of $CP$ asymmetries in $B^{\pm}\to DK^{\pm}$. We present the first measurement of the $CP$-even fraction $F_{+}$ for the decay $D \rightarrow K_{\rm S}^{0}\pi^{+}\pi^{-}\pi^{0}$ which gives $F_{+}$~=~0.238~$\pm$~0.012~$\pm$~0.012. The $F_+$ measurement can be used in a quasi-GLW analysis in which there is no binning of the $D\to K_{\mathrm S}\pi^{+}\pi^{-}\pi^{0}$ phase space, although this does not provide single-mode sensitivity to $\gamma$.  In addition, the measurements of amplitude weighted averages of the cosine and sine of the strong-phase difference between $D^{0}$ and $\bar{D^{0}}$ decaying to the self-conjugate final state of $K_{\rm S}^{0}\pi^{+}\pi^{-}\pi^{0}$ have been performed. This is done in nine regions of the decay phase space binned according to the intermediate resonances present. These results allow a model-independent GGSZ estimation of $\gamma$ from this mode alone. It is estimated that a single-mode uncertainty on $\gamma$ of $\sigma_{\gamma}=4.4^{\circ}$ is achievable with a 50~$\mathrm{ab}^{-1}$ sample of data at Belle II experiment. This could be improved with optimized $c_{i}$ and $s_{i}$ values provided a proper amplitude model is available and a finer binning using a larger sample of quantum-correlated data from the BESIII experiment.

\acknowledgments
We acknowledge the erstwhile CLEO collaboration members for the privilege of using the data for the analysis presented. We would like to thank UK-India Education and Research Initiative for the financial support. We also thank Sam Harnew for fruitful discussions on relations between the bins.


\begin{thebibliography}{99}
 \bibitem{C} N. Cabibbo, {\it Unitary symmetry and leptonic decays}, {\it Phys. Rev. Lett.} {\bf 10} (1963) 531.
 \bibitem{KM} M. Kobayashi and T. Maskawa, {\it CP violation in the renormalizable theory of weak interaction}, {\it Prog. Theor. Phys.} {\bf 49} (1973) 652.
 \bibitem{PDG} Particle Data Group Collaboration, C. Patrignani  et al., {\it Review of Particle Physics}, {\it Chin. Phys. C} {\bf 40} (2016) 100001.
 \bibitem{GLW1} M. Gronau and D. London, {\it How to determine all the angles of the unitarity triangle from $B_{d}\to DK_{\rm S}^{0}$ and $B_{s}^{0}\to D\phi$}, {\it Phys. Lett. B} {\bf 253} (1991) 483. 
 \bibitem{GLW2} M. Gronau and D. Wyler, {\it On determining a weak phase from CP asymmetries in charged $B$ decays}, {\it Phys. Lett. B} {\bf 265} (1991) 172. 
 \bibitem{MNayak} M. Nayak  et al., {\it First determination of the CP content of $D\to \pi^{+}\pi^{-}\pi^{0}$ and $D\to K^{+}K^{-}\pi^{0}$}, {\it Phys. Lett. B} {\bf 740} (2015) 1, [arXiv:1410.3964].
 \bibitem{GGSZ} A. Giri, Yu. Grossman, A. Soffer and J. Zupan, {\it Determining $\gamma$ using $B^{\pm}\to DK^{\pm}$ with multibody D decays}, {\it Phys. Rev. D} {\bf 68} (2003) 054018, [hep-ph/0303187].

 \bibitem{GGSZ2} A. Bondar, {\it Proceedings of BINP special analysis meeting
on Dalitz analysis}, 2002 (unpublished).
 \bibitem{4picisi} S. Harnew et al., {\it Model-independent determination of the strong phase difference between $D^{0}$ and $\bar{D^{0}} \to \pi^{+}\pi^{-}\pi^{+}\pi^{-}$ amplitudes}, [arXiv:1709.03467].
  \bibitem{4pi} S. Malde et al., {\it First determination of the CP content of $D\to \pi^{+}\pi^{-}\pi^{+}\pi^{-}$ and updated determination of the CP contents of $D\to \pi^{+}\pi^{-}\pi^{0}$ and $D\to K^{+}K^{-}\pi^{0}$}, {\it Phys. Lett. B} {\bf 747} (2015) 9, [arXiv:1504.5878].
  
   \bibitem{KsPiPi:EPJC1} A. Bondar and A. Poluektov, {\it Feasibility study of model-independent approach to $\phi_{3}$ measurement using Dalitz plot analysis}, {\it Eur. Phys. J. C} {\bf 47} (2006) 347, [arXiv:hep-ph/0510246].
 \bibitem{KsPiPi:EPJC2} A. Bondar and A. Poluektov, {\it The use of quantum-correlated $D^{0}$ decays for $\phi_{3}$ measurement}, {\it Eur. Phys. J. C} {\bf 55} (2008) 51, [arXiv:0801.0840].
 
 \bibitem{CLEO1} Y. Kubota et al., {\it The CLEO II detector},  {\it Nucl. Instrum. Meth. A} {\bf 320} (1992) 66.
 \bibitem{CLEO2} D. Peterson et al., {\it The CLEO III detector}, {\it Nucl. Instrum. Meth. A} {\bf 478} (2002) 142.
 \bibitem{CLEO3} M. Artuso et al., {\it Construction, pattern recognition and performance of the CLEO III LiF-TEA RICH detector}, {\it Nucl. Instrum. Meth. A} {\bf 502} (2003) 91.
 \bibitem{CLEO4} CLEO-c/CESR-c Taskforces and CLEO-c Collaboration, R.A. Briere et al., {\it CLEO-c and CESR-c: a new frontier of weak and strong interactions}, {\it Cornell
LEPP Report CLNS} Report No. 01/1742 (2001).

 \bibitem{Evtgen} D.J. Lange, {\it The EvtGen particle decay simulation package}, {\it Nucl. Instrum. Meth. A} {\bf 462} (2001) 152.
 \bibitem{Geant} R. Brun et al., {\it GEANT 3.21}, {\it CERN Program Library Long Writeup
W5013} (unpublished).

\bibitem{PHOTOS} E. Barberio and Z. Was, {\it PHOTOS - a universal Monte Carlo for QED radioactive corrections: version 2.0}, {\it Comput. Phys. Commun.} {\bf 79} (1994) 291.

 \bibitem{MissMass} CLEO Collaboration, N. Lowrey  et al., {\it Determination of the $D^{0}\to K^{-}\pi^{+}\pi^{0}$ and $D^{0}\to K^{-}\pi^{+}\pi^{-}\pi^{+}$ coherence factors and average strong-phase differences using quantum-correlated measurements}, {\it Phys. Rev. D} {\bf 80} (2009) 031105, [arXiv:0903.4853].
 
 \bibitem{ARGUS} ARGUS Collaboration, H. Albrecht et al., {\it Search for hadronic $b\to u$ decays },  {\it Phys. Lett. B} {\bf 241} (1990) 278.
 \bibitem{CB} T. Skwarnicki, {\it A study of the radiative cascade transitions between the $\Upsilon$ and $\Upsilon'$ resonances}, {\it Ph.D Thesis (Appendix E)}, DESY F31-86-02 (1986).
 
 \bibitem{Dmix} Heavy Flavor Averaging Group Collaboration, Y. Amhis et al., {\it Averages of b-hadron, c-hadron and $\tau$-lepton properties as of November 2016}, [arXiv:1612.07233].
 \bibitem{KsPiPi} CLEO Collaboration, J. Libby  et al., {\it Model-independent determination of the strong-phase difference between $D^{0}$ and $\bar{D^{0}} \to K_{\rm S,L}^{0}h^{+}h^{-}~(h = \pi,K)$ and its impact on the measurement of the CKM angle $\gamma/\phi_{3}$}, {\it Phys. Rev. D} {\bf 82} (2010) 112006, [
arXiv:1010.2817].
 \bibitem{BaBar} BaBar Collaboration, B. Aubert et al., {\it Improved measurement of the CKM angle $\gamma$ in $B^{\mp}\to D^{(*)}K^{(*\mp)}$ decays with a Dalitz plot analysis of D decays to $K_{\rm S}^{0}\pi^{+}\pi^{-}$ and $K_{\rm S}K^{+}K^{-}$}, {\it Phys. Rev. D} {\bf 78} (2008) 034023, [arXiv:0804.2089].

 \bibitem{Belle-GGSZ} Belle Collaboration, H. Aihara et al., {\it First measurement of $\phi_{3}$ with a model-independent Dalitz plot analysis of $B^{\pm}\to DK^{\pm}$, $D\to K_{\rm S}^{0}\pi^{+}\pi^{-}$ decay}, {\it Phys. Rev. D} {\bf 85} (2012) 112014, [arXiv:1204.6561].

\bibitem{Prasanth} Belle Collaboration, K. Prasanth et al., {\it First measurement of $T$-odd moments in $D^{0}\to K_{\rm S}^{0}\pi^{+}\pi^{-}\pi^{0}$ decays}, {\it Phys. Rev. D} {\bf 95} (2017) 091101(R), [arXiv:1703.05721].
 
\bibitem{BelleD2KsPiPi} Belle Collaboration, T. Peng et al., {\it Measurement of $D^{0}-\bar{D^{0}}$ mixing and search for indirect CP violation using $D^{0}\to K_{\rm S}^{0}\pi^{+}\pi^{-}$ decays}, {\it Phys. Rev. D} {\bf 89} (2014) 091103(R), [arXiv:1404.2412].

\bibitem{CKMfitter} http://ckmfitter.in2p3.fr
 
 \bibitem{b2tip} E. Kou, P. Urquijo, The Belle II collaboration, and The B2TiP theory community, {\it The Belle II Physics Book}, in preparation.

\end{thebibliography}
\end{document}